\documentclass[pra,superscriptaddress]{revtex4}
\usepackage{graphicx}
\usepackage{url}
\usepackage{amsfonts,amsmath,amssymb,float}
\usepackage{bm,dsfont}
\usepackage{color}
\usepackage{bbm}
\usepackage{braket}
\usepackage{longtable}
\usepackage[dvips]{epsfig}
\usepackage{amsmath,amssymb,lscape,float}
\usepackage{hyperref}
\usepackage{breakurl}
\usepackage{listings}
\usepackage{etoolbox}

\begin{document}
\title{Digital Quantum Simulation of Scalar Yukawa Coupling} 

\author{Thierry N. Kaldenbach} 
\affiliation{Institut f\"{u}r Angewandte Physik, Technical University of Darmstadt, 
D-64289 Darmstadt, Germany}
\affiliation{Fraunhofer-Institut f\"{u}r Graphische Datenverarbeitung IGD, D-64283 Darmstadt, Germany}

\author{Matthias Heller}
\affiliation{Fraunhofer-Institut f\"{u}r Graphische Datenverarbeitung IGD, D-64283 Darmstadt, Germany}
\affiliation{Interactive Graphics Systems Group, Technical University of Darmstadt, D-64283, Darmstadt, Germany}

\author{Gernot Alber}
\affiliation{Institut f\"{u}r Angewandte Physik, Technical University of Darmstadt, 
D-64289 Darmstadt, Germany}

\author{Vladimir M. Stojanovi\'c}
\email{vladimir.stojanovic@physik.tu-darmstadt.de}
\affiliation{Institut f\"{u}r Angewandte Physik, Technical University of Darmstadt, 
D-64289 Darmstadt, Germany}

\date{\today}

\begin{abstract}
Motivated by the revitalized interest in the digital simulation of medium- and high-energy physics 
phenomena, we investigate the dynamics following a Yukawa-interaction quench on IBM Q. Adopting the 
zero-dimensional version of the scalar Yukawa-coupling model as our point of departure, we design low-depth 
quantum circuits emulating its dynamics with up to three bosons. In the one-boson case we demonstrate 
circuit compression, i.e., a constant-depth circuit containing only two controlled-NOT (CNOT) gates. In the 
more complex three-boson case, we design a circuit in which one Trotter step 
entails $8$ CNOTs. Using an analogy with the travelling-salesman problem, we also provide a CNOT-cost estimate 
for higher boson-number truncations. Based on these circuits, we quantify the system dynamics by evaluating the expected boson number at an arbitrary time after the quench and the survival probability of the initial vacuum state (the Loschmidt echo). We also utilize these circuits to drive adiabatic transitions and compute the energies of the ground- and first-excited states of the considered model. Finally, 
through error mitigation -- i.e, zero-noise extrapolation -- we demonstrate a good agreement of our results with a numerically-exact classical benchmark.
\end{abstract}

\maketitle

\section{Introduction}
The reinvigorated interest in {\em digital quantum simulation} (DQS)~\cite{Lloyd:96,Zalka:98,Georgescu+:14} 
has chiefly been motivated by the tantalizing recent progress in the development of quantum hardware based on superconducting- \cite{Wendin:17},
trapped-ion-~\cite{Bruzewicz+:19}, or neutral-atom systems~\cite{MorgadoWhitlock:21}. 
This recent flurry of research activity in DQS was preceded by a large body of investigation pertaining to analog quantum simulators of many-body systems~\cite{Georgescu+:14}. Both of these research strands have in large part been inspired by the pioneering assertion of Feynman that 
a generic quantum system could efficiently be simulated using a device whose operation is also governed by quantum-mechanical 
laws~\cite{feynmansim}. Admittedly, the field of DQS has heretofore been dominated by the development and application 
of quantum algorithms for simulating purely fermionic systems~\cite{Abrams+Lloyd:9799,Somma+:02,Bravyi+Kitaev:02,
Whitfield+:11,Raeisi+:12,Wecker+:15,Barends+:15,Babbush+:18,Reiner+:18,Jiang+:18,McArdle+:20}; this is related to the fact that -- owing to the Pauli exclusion principle -- simulations thereof typically require modest quantum-hardware resources. 

Unlike their purely fermionic counterparts, quantum many-body systems involving 
bosonic constituents (e.g. coupled boson-boson~\cite{Hofer+:12} or boson-fermion~\cite{Macridin+:18} systems) have infinite-dimensional Hilbert spaces that ought to be truncated if one aims to simulate such systems on either classical or quantum computers.
Due to the inherently nontrivial problem of encoding bosonic states on qubit registers, the DQS
of bosonic systems has heretofore received comparatively modest attention~\cite{Macridin+:18,Macridin+:22,Miessen+:21}. In particular, while many analog simulators of {\em coupled boson-fermion} models~\cite{Stojanovic:20} have been proposed~\cite{Stojanovic+:12,Mei+:13,Stojanovic+:14,Stojanovic+Salom:19,StojanovicPRL:20,Stojanovic:21,
Nauth+Stojanovic:23}, only a handful of such models have as yet been addressed in the DQS context~\cite{Macridin+:18,Tong+:22}. 

Motivated primarily by the dearth of works pertaining 
to the DQS of coupled boson-fermion models, in this paper we present a 
DQS of the nonequilibrium dynamics following a quench~\cite{Heyl:18}
of the scalar Yukawa interaction~\cite{PeskinSchroederBOOK}. While 
our principal goal here is to demonstrate a DQS of a coupled boson-fermion
model of physical interest -- demonstrating in the process efficient schemes for a controlled truncation of (infinite-dimensional) bosonic Hilbert spaces -- the present work also fits seamlessly into
the emerging research area that pertains to the DQS of models 
from medium- and high-energy physics~\cite{Martinez+:16,Holland+:20,Tong+:22,Kreshchuk+:22,
Nguyen+:22,Kico+:22}. 

Yukawa-type interactions -- first proposed for describing the 
nucleon-nucleon interaction mediated by pions~\cite{Yukawa:35} -- involve fermion-, antifermion-, and boson degrees of 
freedom. Due to limitations of current noisy intermediate-scale quantum hardware~\cite{PreskillNISQ:18}, 
here we investigate the low-energy limit of the scalar Yukawa coupling on a single lattice site, with low truncation numbers
for the real scalar (boson) field. We show that, despite its inherent simplicity, the resulting zero-dimensional model shows nontrivial quantum dynamics. While the simplified, zero-dimensional version of the scalar Yukawa-coupling model cannot be expected to bear out all the relevant features of the actual physical phenomenon, our aim here -- i.e., the second major goal of this work -- is to use the DQS of this simple model as a testbed for state-of-the-art IBM Quantum (IBM Q) hardware~\cite{ibmq}.

By first making use of the local charge conservation in the system at hand, which permits a single-qubit encoding of the fermion-antifermion sector of the problem, we design low-depth quantum circuits that emulate the system dynamics with up to three bosons. We subsequently implement those circuits on IBM Q hardware~\cite{ibmq}, where CNOT is the representative two-qubit gate. We characterize the system dynamics for the initial vacuum state by computing the expected boson numbers and the survival probability of the initial state (the Loschmidt echo)~\cite{Peres:85} at an arbitrary time after the Yukawa-interaction quench. In addition, we construct circuits for calculating the energies of the ground- and first-excited states using adiabatic state preparation~\cite{Fock,Albash+Lidar:18}. We also provide benchmarks for the obtained results -- in both the post-quench-dynamics-
and adiabatic-state-preparation parts of this work -- by carrying out numerically-exact evaluations of the aforementioned quantities on a classical computer. Finally, we perform error-mitigation in the form of digital {\em zero-noise extrapolation}~\cite{Temme+:17, Li+Benjamin:17}, showing in this manner that our DQS results are in a good agreement with the numerically-exact classical benchmarks.

Using sophisticated circuit-optimization techniques, in the one-boson case we demonstrate circuit compression, i.e. design a constant-depth circuit that contains only two CNOT gates (regardless of the total simulation time). In other words, in this case we find a 
much more efficient realization than the conventional one in which the circuit depth scales linearly with the number 
of Trotter steps. Such circuit compression is only possible for certain special types of the system (coupled-qubit) 
Hamiltonian and have quite recently been discussed in the context of the transverse-field Ising-~\cite{Stojanovic+Nauth:22,
Stojanovic+Nauth:23} 
and $XY$~\cite{StojanovicPRL:20} models by means of Lie-algebraic methods~\cite{Kokcu+:22,Peng+:22}. 

In the more nontrivial three-boson case -- which in our state-encoding scheme corresponds to a three-qubit system -- a compression to 
a constant-depth circuit is no longer possible. We show that in this case it proves beneficial to make use of the second-order Trotter-Suzuki product formula~\cite{Suzuki:76,Hatano+Suzuki:05}, which entails 
a symmetrized Trotter step; in this way, we design a circuit in which one Trotter step entails $8$ CNOTs. This last number is far below the maximal 
CNOT-cost of a generic three-qubit gate ($14$) ~\cite{Shende+:04,Stojanovic:19}.

We also address cases with larger boson-number truncations by providing the CNOT-cost estimate for the corresponding quantum circuits. We do so based on an analogy of our circuit optimization with the \textit{travelling-salesman problem} (TSP)~\cite{SkienaBOOK:12}, adapting an idea proposed in Ref.~\cite{Gui+:20}. 
To this end, we make use of the exact solution of the TSP based on the Bellman-Held-Karp dynamic-programming 
algorithm~\cite{Bellman:62,Held+Karp:62}, as well as the Christofides algorithm, a polynomial-time heuristic that approximately solves the TSP on a metric graph~\cite{Christofides:76}. Being polynomial in character, the latter can be utilized for much larger boson-number truncations than the exact solution.  

The remainder of this paper is organized as follows. In Section~\ref{SystemHamiltonian} we introduce the relevant fermion-boson system and its underlying Hamiltonian. Section~\ref{DQSandEncodeBF} starts with a short introduction into our scheme for describing the system dynamics after an interaction quench via DQS, followed by a brief recapitulation of our adopted approaches for encoding fermion- and boson states. At the end of this section, we also address the issue of the controllable truncation of the bosonic Hilbert space. Section~\ref{circ_synth} starts with the design of low-depth quantum circuits emulating the dynamics of the system, first in the two-qubit- and then in the three-qubit case. This is followed by an analysis of the
CNOT-cost for higher boson-number truncations. Finally, this section ends with a brief review of the basic aspects of the zero-noise extrapolation technique of error mitigation. The principal results obtained for the system dynamics on IBM Q
are presented in Section~\ref{ResDisc}. We summarize the main findings and conclude the paper in Section~\ref{SummConcl}. Some cumbersome derivations, as well as relevant mathematical and computational details, are relegated to Appendices~\ref{DeriveHamiltonian}~-~\ref{CNOTcostBound}. 

\section{System and Hamiltonian} \label{SystemHamiltonian}
To set the stage for further considerations, we first introduce the system at hand and its 
fermion-boson Hamiltonian. We start  by introducing the scalar Yukawa-coupling model (Section~\ref{scalarYukawa}) and then specialize to its single-site 
(zero-dimensional) version (Section~\ref{SSmodel}).

\subsection{Scalar Yukawa coupling} \label{scalarYukawa}
The Yukawa-interaction mechanism was originally introduced to describe the nuclear force between 
nucleons mediated by pions~\cite{Yukawa:35}. It later attracted interest in the context of the Standard 
Model of particle physics, where it describes the coupling between the Higgs field and massless quark 
and lepton fields (i.e.~the fundamental fermion fields); the latter fermion fields acquire masses, 
via the Higgs mechanism, after electroweak symmetry breaking~\cite{PeskinSchroederBOOK}. Interaction mechanism somewhat analogous to Yukawa coupling are also of 
relevance in the context of strongly-correlated 
condensed-matter systems~\cite{RanningerBFM:95}.

In the following, we will be interested in the Yukawa-type interaction of a real scalar field $\phi$ and a 
Dirac field $\psi$, described by the Hamiltonian~\cite{PeskinSchroederBOOK}
\begin{equation}\label{eq:HIntScalar}
H_\text{int} = g \int d^3\mathbf{x} \:\bar\psi\psi\phi \:,\\
\end{equation}
where $g$ is the dimensionless coupling strength. While pions, which mediate the nucleon-nucleon interaction,
are pseudoscalar mesons, we will assume a scalar interaction for the sake of simplicity. In our subsequent 
discussion, we adopt natural units, where $\hbar=c=1$.

To perform a DQS of the real-time dynamics of scalar Yukawa coupling, we have to discretize 
the theory and select a convenient basis. While approaches that are directly based on the algebraic properties 
of the field operators do exist, in what follows we make use of an expansion in terms of creation and annihilation 
operators. We further select the momentum eigenstates $|\textbf{p}\rangle$ as our preferred basis, as these these states are at the same time eigenstates of the non-interacting (free) part of the Hamiltonian of the 
system. Using the definition of the Fourier transform $\phi(\textbf{x}) = {(2\pi)}^{-3}\int
d^3\mathbf{p}\:e^{i \textbf{p}\cdot \textbf{x}}\phi(\textbf{p})$, one may rewrite the interaction Hamiltonian as 
\begin{eqnarray}
H_\text{int} &=& \frac{g}{2} \int \frac{d^3\mathbf{p}}{(2\pi)^3} \int\frac{d^3\mathbf{p'}}
{(2\pi)^3} \left[\bar\psi(\textbf{p})\psi(\textbf{p}')\phi(\textbf{p}-\textbf{p}') +\text{H.c.}\right],
\label{eq:HIntScalarMomentum}
\end{eqnarray}
with the boson- and fermion/antifermion field operators in momentum space being respectively defined as~\cite{PeskinSchroederBOOK} 
\begin{equation}
\begin{split}
 \phi(\textbf{p}) &= \frac{1}{\sqrt{2\omega_\textbf{p}}}\left(b_\textbf{p}+ b_{-\textbf{p}}^\dagger\right), \\
 \psi(\textbf{p}) &= \frac{1}{\sqrt{2 \Omega_\textbf{p}}} \sum_s \left[a_\textbf{p}^s  u^s(\textbf{p}) 
+ c_{-\textbf{p}}^{s\dagger}  v^s(-\textbf{p})\right]\:,
\end{split}
\label{eq:FieldOpsMomentum}
\end{equation}
with the relativistic dispersion relations $\omega_\textbf{p}^2 = m^2+\textbf{p}^2$ (for bosons) and $\Omega_\textbf{p}^2 = 
M^2 + \textbf{p}^2$ (for fermions)~\cite{PeskinSchroederBOOK}. By discretizing the integrals in Equation~\eqref{eq:HIntScalarMomentum} 
and using Equation~\eqref{eq:FieldOpsMomentum}, one obtains the momentum-space version
of $H_\text{int}$ in terms of creation/annihilation operators
\begin{eqnarray}\label{fullHint}
H_\text{int} &=& \frac{ga_{l}^{3/2}}{2} \sum_{\textbf{p}, \textbf{p}'} \sum_{r, s} \Big\{ 
\frac{1}{\sqrt{8 \Omega_\textbf{p}\Omega_{\textbf{p}'}\omega_{\textbf{p}-\textbf{p}'}}} 
\left(b_{\textbf{p}-\textbf{p}'}+ b_{\textbf{p}'-\textbf{p}}^\dagger\right)\nonumber \\
&\times& \big[a_\textbf{p}^{r\dagger} \bar u^r(\textbf{p}) + c_{-\textbf{p}}^{r} \bar v^r(-\textbf{p})\big]\\
&\times& \big[a_{\textbf{p}'}^s  u^s(\textbf{p}') + c_{-\textbf{p}'}^{s\dagger}  v^s(-\textbf{p}')\big]
+\text{H.c.} \Big\}\:, \nonumber
\end{eqnarray}
where $a_{l}$ is the lattice spacing.

Having discussed the interacting part of the full Hamiltonian
of the system, it remains to write down its free part $H_0$. The latter
consists of the free Hamiltonians of the Dirac- and Klein-Gordon fields in momentum space, i.e. $H_0=H_\text{Dirac}+H_\text{KG}$, where
\begin{equation} \label{EqFreeHamiltonian}
\begin{split}
H_\text{Dirac} &= \int \frac{d^3p}{(2\pi)^3}\:\Omega_\textbf{p} \sum_{s} 
\left(a_\textbf{p}^{s\dagger} a_\textbf{p}^{s} + c_\textbf{p}^{s\dagger} c_\textbf{p}^{s}\right) \:,\\
H_\text{KG} &= \int \frac{d^3p}{(2\pi)^3}\:\omega_\textbf{p} b_\textbf{p}^{\dagger} b_\textbf{p} \:.
\end{split}
\end{equation}

\subsection{Single-site model} \label{SSmodel}
Due to technological limitations of current quantum hardware, it appears prudent to start investigating scalar 
Yukawa interaction on a single lattice site with low truncation numbers for the real scalar. While for large 
lattices the gate cost of the time evolution scales more favourably in real space, the momentum-space representation is 
still suitable for small lattices -- especially in the extreme case of a single-site model. The single grid point 
in momentum space corresponds to a momentum below a certain small cutoff value for all particles, implying that 
all particles can be considered to be approximately at rest. While zero-dimensional, single-site models of this type 
are widely used in the context of the DQS of high/medium energy models~\cite{Ciavarella:20,Farrell+:22}, it is 
obvious that they cannot be expected to yield accurate results for any relevant physical observable. However, such 
``toy models'' can still display nontrivial quantum dynamics and can thus be seen as a useful first step towards 
more complex simulations that should be achievable in the not-too-distant future.

As derived in Appendix~\ref{DeriveHamiltonian}, the effective interaction Hamiltonian of our single-site 
model is given by
\begin{equation} \label{ssHint}
H_\text{int} = \frac{\eta}{2}\left(a^{\dagger} a+c^{\dagger} c -1\right)
\left(b+ b^\dagger\right)\:,
\end{equation}
with $\eta\equiv 4mg\beta^{3/2}$ being the effective coupling strength. At the same time, the single-site 
version of the free Hamiltonian $H_0 = H_\text{Dirac} + H_\text{Klein-Gordon}$ is given by 
\begin{equation} \label{ssH0}
H_0 = M (a^{\dagger} a + c^{\dagger} c) + m b^\dagger b \:.
\end{equation}
The sum of the Hamiltonians in Equations~\eqref{ssHint} and \eqref{ssH0} -- i.e. the total Hamiltonian 
$H_\text{tot}=H_0+H_\text{int}$ -- will be our point of departure for simulating the dynamics following a quench of the scalar Yukawa coupling in what follows.

\section{DQS of Hamiltonian dynamics and encoding fermion/boson states on qubits}   \label{DQSandEncodeBF}
To facilitate the discussion of the DQS of the Yukawa-coupled system in the remainder of this paper, we begin with some general aspect of DQS and quantum-circuit synthesis (Section~\ref{DQSforQuench}). We then recapitulate the basic aspects of encoding the relevant particle states on qubits, both for fermions (Section~\ref{EncodeFermions}) and bosons (Section~\ref{EncodeBosons}). Finally, we discuss the truncation of the boson Hilbert space (Section~\ref{sec:Truncation}).

\subsection{Simulating the dynamics after an interaction quench}   \label{DQSforQuench}
For a quantum system described by a Hamiltonian $H_{\textrm{sys}}$, the Loschmidt amplitude~\cite{Heyl:18}
\begin{equation}\label{LoschmidtAmp}
\mathcal{G}(t)=\langle\psi_{t=0}|\:e^{-i H_{\textrm{sys}}t}\:|\psi_{t=0}\rangle
\end{equation}
quantifies the deviation of the time-evolved state $|\psi_{t}\rangle\equiv\:e^{-iH_{\textrm{sys}}t}
\:|\psi_{t=0}\rangle$ from the initial state $|\psi_{t=0}\rangle$. The Loschmidt echo~\cite{Peres:85} $\mathcal{L}(t)\equiv|\mathcal{G}(t)|^2$, the survival probability at time $t$ of this initial state, represents the most widely used quantity in describing dynamics following an interaction quench. 
In order to evaluate this last quantity, as well as 
the expected boson number, at time $t$ after a 
Yukawa-interaction quench, one first evaluates the time-evolution operator of the system $U(t)\equiv e^{-iH_{\textrm{tot}}t}$ by representing it in the form of a quantum circuit.

The unitary time-evolution operator $U(t)$ corresponding to the Hamiltonian 
$H(t)$ satisfies the time-dependent Schr\"{o}dinger equation in the operator form $\partial U/\partial 
t = -iH(t)U(t)$, with the initial condition $U(t=0)=\mathbbm{1}$. We will hereafter adopt a 
time-independent Hamiltonian $H(t) = H_{\textrm{sys}}$. By discretizing time into $n$ steps of duration $\Delta t$, such that $t\equiv n\Delta t$ is the total evolution time, the time-evolution operator can be expressed as 
\begin{equation}
 U(t) = (e^{-iH_{\textrm{sys}}\Delta t})^{n} \:.
\end{equation}
The inherent tradeoff pertaining to such a decomposition is that a smaller time step corresponds to a longer 
circuit. The last expression for $U(t)$ is typically approximated using first-order Trotter-Suzuki-type 
decomposition~\cite{Trotter:59,Suzuki:76}. 
For a time-independent Hamiltonian $H_{\textrm{sys}}=\sum_l H_l$ the latter approximates $\exp(-i H_{\textrm{sys}}t)$ by 
$(\prod_l e^{-i H_l t/n})^{n}$, with the corresponding error being upper bounded by $\mathcal{O}(Nt^2/n)$, 
where $N$ is the number of qubits~\cite{Lloyd:96}. 

In the DQS context, time-evolution operators are represented through quantum circuits that emulate the system dynamics and are decomposed into single-qubit and two-qubit gates. Typical single-qubit gates include
the Pauli gates~\cite{NielsenChuangBook} 
\begin{equation}
X=\begin{bmatrix} 
    0 & 1\\
    1 & 0\\
\end{bmatrix} 
\:,\:Y=  \begin{bmatrix} 
    0 & -i\\
    i & 0\\
\end{bmatrix}
\:,\:  
Z= \begin{bmatrix} 
    1 & 0\\
    0 & -1\\
\end{bmatrix}\:,
\end{equation}
the single-qubit rotation gates 
\begin{equation}
 R_x(\theta)=e^{-\frac{i}{2}\theta X}\:,\:R_y(\theta)=e^{-\frac{i}{2}\theta Y}
 \:,\: R_z(\theta)=e^{-\frac{i}{2}\theta Z}\:,
\end{equation}
as well as the Hadamard- ($H$), phase- ($S$) and $\pi/8$ ($T$) gates
\begin{equation}
H=\frac{1}{\sqrt{2}}\begin{bmatrix} 
  1 & 1\\
  1 & -1\\
\end{bmatrix} 
\:,\:
S=\begin{bmatrix} 
  1 & 0\\
  0 & i\\
\end{bmatrix}
\:,\:
T=\begin{bmatrix} 
  1 & 0\\
  0 & e^{i\pi/4}\\
\end{bmatrix} \:.
\end{equation}
The essential two-qubit gate is CNOT, whose matrix representation is 
\begin{equation}
\textrm{CNOT} = \begin{bmatrix} 
1 & 0 & 0 & 0\\
0 & 1 & 0 & 0\\
0 & 0 & 0 & 1\\
0 & 0 & 1 & 0
\end{bmatrix}\:.
\end{equation}  

Simulating the dynamics of a quantum many-body system on a classical computer is a nontrivial problem due to the large Hilbert-space size that necessitates the allotment of a large amount of memory. The task 
of designing a circuit that emulates these same dynamics on a quantum computer~\cite{NielsenChuangBook} is exponentially hard, which is intimately related to exponential growth of a generic Hilbert space with the system size; e.g., in the case of interacting fermion systems this is known to be an NP-hard problem~\cite{Abrams+Lloyd:98,ZhengSciRep:21}. What further complicates this task is the fact that near-term quantum hardware is both very limited in size and noisy~\cite{PreskillNISQ:18}. Thus, 
the optimization of the quantum circuit emulating the dynamics of a relevant system -- i.e. reducing its depth in order to obtain as shallow a circuit as possible -- is a task of paramount importance for DQS.
  
\subsection{Fermion-state encoding}   \label{EncodeFermions}
In the familiar Jordan-Wigner (JW) encoding~\cite{Jordan+Wigner:28} of fermion states, due to the Pauli exclusion 
principle, the fermionic occupation numbers are restricted to the set $\{0, 1\}$. This allows for a one-to-one mapping 
from the occupation-number basis to the computational basis - a qubit in the state $\ket{0}$($\ket{1}$) corresponds to an empty (occupied) fermionic orbital. Therefore,
\begin{equation} \label{fermstates}
\ket{f_{n-1} \dots f_1 f_0} \to \ket{q_{n-1}} \otimes \dots \otimes 
\ket{q_1} \otimes \ket{q_0} \:,
\end{equation}
where $f_j$ is the occupation number of orbital $j$ ($j=0,\ldots,n-1$) and $\ket{q_j}$ the corresponding qubit
state [$f_j = q_j \in \{0, 1\}$]. In Equation~\eqref{fermstates} the rightmost 
single-qubit state corresponds to the qubit $0$, 
a convention that will hereafter also apply to operators.

The JW-type encoding requires $O(n)$ qubit operations to represent a fermionic 
operator, where $n$ is the number of fermionic orbitals (or, in the problem at hand, momentum states of 
spin-up and spin-down particles/antiparticles). While the alternative, Bravyi-Kitaev encoding~\cite{Bravyi+Kitaev:02,Seeley+:12} 
reduces this cost to $O(\log\:n)$, in what follows we will rely on the JW encoding as we will be concerned with
Hamiltonians that are diagonal on the fermionic subspace in the occupation-
number basis.

The JW qubit mappings of fermionic operators are given by 
\begin{eqnarray}\label{eq:JWMapping}
c_j^\dagger &\equiv& \frac{1}{2}\:(X_j-iY_j)\otimes Z_{j-1} \otimes \dots \otimes Z_1 
\otimes Z_0 \:,\nonumber \\
c_j &\equiv& \frac{1}{2}\:(X_j+iY_j)\otimes Z_{j-1} \otimes \dots \otimes Z_1 \otimes Z_0 \:.
\end{eqnarray}
Here the operators $(X_j\pm iY_j)/2$ change the occupation numbers of target spin orbital, while the action 
of the string of $Z$ operators amounts to computing the parity of the state. 

\subsection{Boson-state encoding}   \label{EncodeBosons}
Bosonic creation/annihilation operators satisfy the commutation relation $[b_i, b_j^\dagger] = \delta_{ij}$ and, by contrast to fermions, commute on different sites. The main challenge in encoding bosonic states pertains to their unbounded occupation number, which forces us to impose an occupation-number cutoff $\Lambda$. The 
encoding of a bosonic Fock state with cutoff $\Lambda$ requires $O(n\log\Lambda)$ qubits, where $n$ is 
the number of bosonic modes. This scaling is achieved by representing the occupation numbers of the basis 
states of the truncated Fock space as binary strings and mapping the digits to individual qubits~\cite{Macridin+:18}. 
For a Fock state $\ket{k}$, where $k$ is an integer whose binary decomposition reads
\begin{equation} \label{BinaryRepresent}
k = \sum_{j=0}^{N-1} q_j(k)\:2^j \quad (\:q_j\in\{0, 1\}\:) \:,
\end{equation}
the mapping is given by
\begin{equation}
\ket{k} \to \ket{q_{N-1}} \otimes \dots \otimes 
\ket{q_1} \otimes \ket{q_0} \:. \label{eq:BosonMapping}
\end{equation}

The mapping of the truncated creation/annihilation operators to their pseudospin--$1/2$ (qubit) counterparts
is then achieved by finding their corresponding Pauli-basis representations for arbitary truncations. For 
$\Lambda = 2^N -1$, a bosonic creation operator can be represented as
\begin{equation}\label{eq:BosonCreationExplicit}
 b^\dagger = \left(\frac{1}{2}\right)^N \sum_{k=1}^{\Lambda}\sqrt{k}
 \: \bigotimes_{j=0}^{N-1}F_{j,k} \:,
\end{equation}
where $F_{j,k}$ is an operator defined as (for a detailed derivation, see Appendix~\ref{BosonMapping})
\begin{equation}\label{eq:DefineF}
F_{j,k}=\begin{cases}
I_j + (-1)^{q_j(k)} Z_j & \textrm{if} \:\exists\ m<j:q_m(k)=1\\ 
X_j + (-1)^{q_j(k)} i Y_j & \textrm{if}\: \forall\ m<j: q_m(k)=0\ \\
\end{cases}\:.
\end{equation}
When taking the expectation value $\braket{b^\dagger b}$, we simply weight the probability
that the $k$-th qubit is in state $\ket{1}$ by a factor of $2^{k}$. 

Another often used operator is the particle-number operator $b^\dagger b$, which in the 
Pauli basis is given by
\begin{equation}\label{eq:BosonPartNumber}
b^\dagger b = \frac{1}{2}\sum_{j=0}^{N-1} 2^j (I-Z)_j \:.
\end{equation}

\subsection{Truncation of the boson Hilbert space} \label{sec:Truncation}

\begin{figure}[b!]
    \centering
\includegraphics[width=0.75\columnwidth]{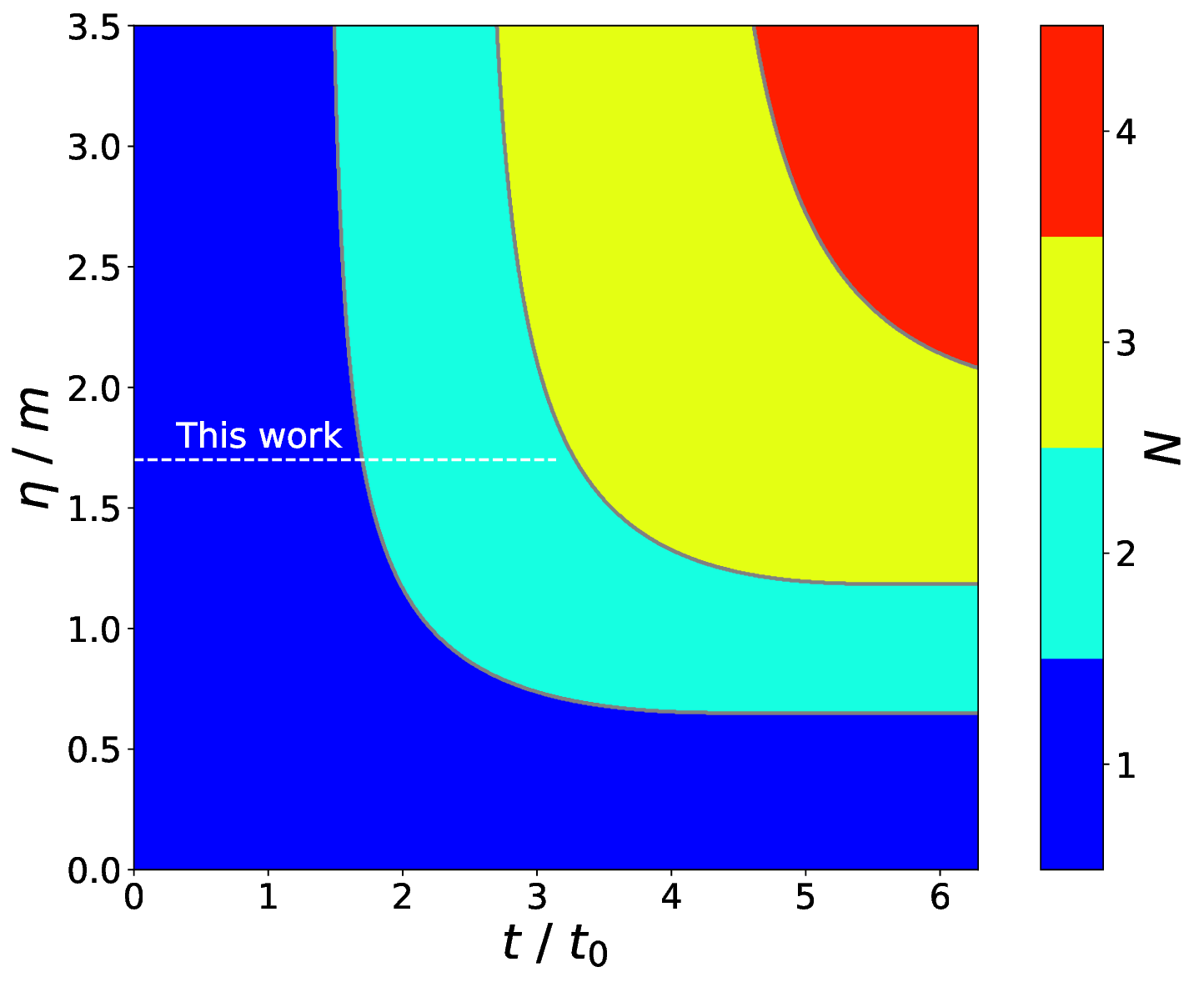}
    \caption{Required number $N$ of qubits as a function of dimensionless time $t/t_0$ and the coupling-to-mass ratio $\eta/m$, where $t_0\equiv 1/\sqrt{m^2 + \eta^2}$. To ensure that the target fidelity of 90\% $(\epsilon = 0.1)$ is attained, we can simulate the system using two bosonic qubits up to a time $t\approx 3\, t_0$.}
    \label{fig:TruncationMap}
\end{figure}

As discussed in Section~\ref{EncodeBosons}, a truncation of the (infinite dimensional) boson Hilbert space is required in order to perform a DQS with a finite number of qubits. The ideal truncation depends non-trivially on the initial state of the system $\ket{\psi_{t=0}}$, the coupling strength $\eta$, and the total simulation time $t$. Throughout this work, we focus entirely on interaction quenches starting from the vacuum state as the initial state of the system. This obviates the need to consider the dependence of the system dynamics following a quench on the initial state. While theoretical estimations of the truncation required to keep the truncation error below a certain threshold are known \cite{Tong+:22}, we will rely on classical simulations of our few-qubit quantum systems in order to determine the ideal qubit requirements. 

Our algorithm works as follows:
\begin{enumerate}
    \item Define a high resolution grid of the coupling strength and the time in the regime of interest. 
    \item For each coupling strength, select an initial truncation with $N=1$ qubits ($\Lambda=1$).
    \item  Simulate a Hamiltonian with a truncation using $N$ and $N+1$ qubits on the time grid.
    \item Compute the fidelity between both simulations for each point on the time grid. This is achieved by extending the lower dimensional state vector into the higher dimensional Hilbert space using the zero state $\ket{\psi_N} \to \ket{0} \otimes \ket{\psi_N}$.
    \item If the fidelity falls below a certain threshold $1-\epsilon$ at some point in time, the selected truncation is no longer appropriate. Then increase the qubit number by one $(N\to N+1)$, save the qubit number along with the time point, and repeat step 3, 4 and 5 until the desired fidelity is reached. 
\end{enumerate}

\begin{figure}[b!]
    \centering
\includegraphics[width=0.75\columnwidth]{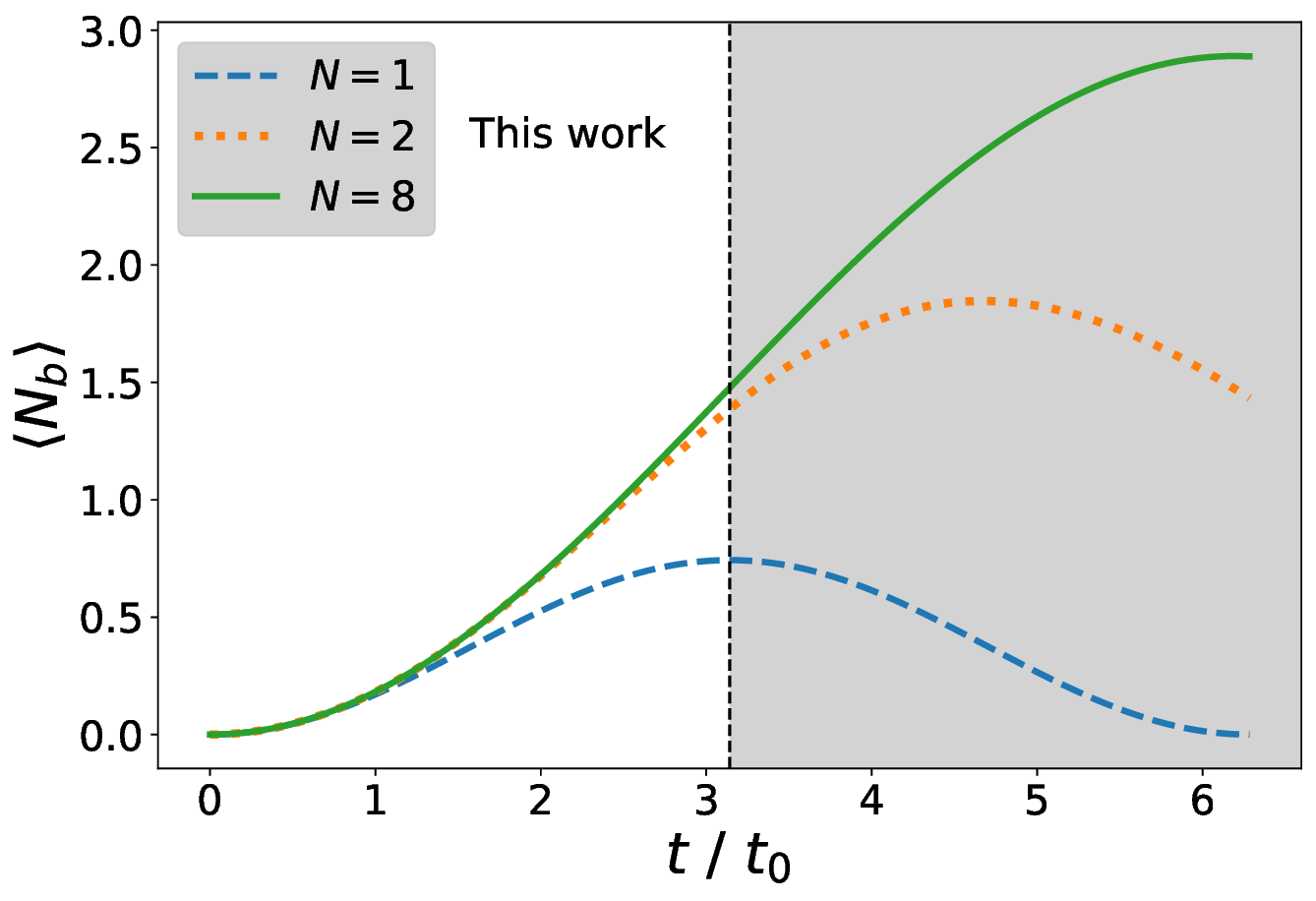}
    \caption{The bosonic occupation number $\braket{N_b}$ as a function of dimensionless time $t/t_0$ for a coupling-to-mass ratio of $\eta/m=1.7$. Note, that the simulation with $N=2$ shows a good agreement with the $N=8$ simulation  up to a time $t\approx 3\, t_0$. The strong deviations for longer times indicate the need for a larger truncation, in agreement with the results illustrated by Figure~\ref{fig:TruncationMap}.}
    \label{fig:BosonClassicalSim}
\end{figure}

\begin{figure}[t!]
    \centering
\includegraphics[width=0.75\columnwidth]{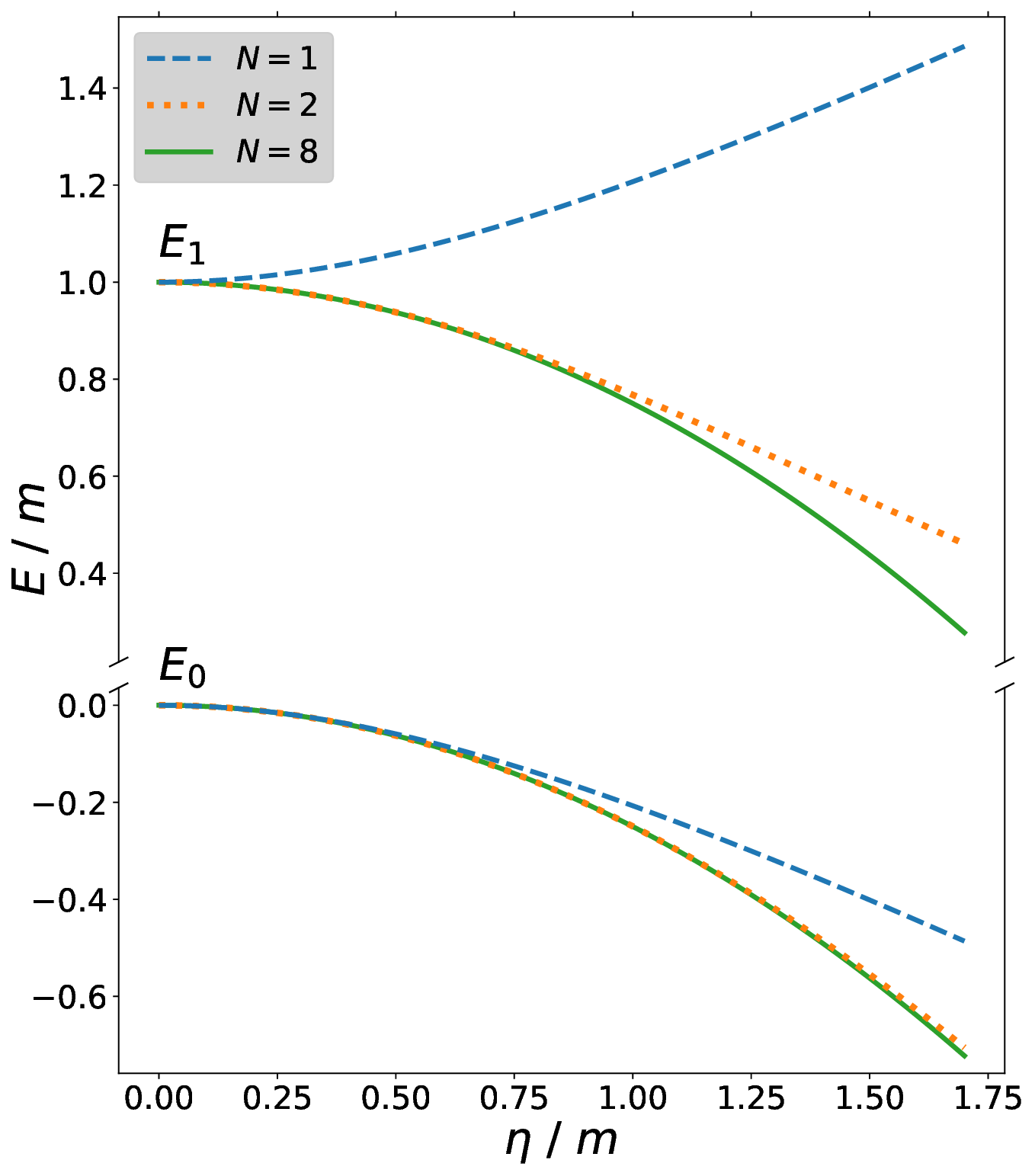}
    \caption{Energy-to-mass ratio $E/m$ of the ground state and first excited state as a function of the coupling-to-mass ratio $\eta/m$.
    A truncation with $N=1$ is insufficient to approximate the first excited state even at low coupling strengths. The truncation with $N=3$ shows excellent agreement with the ground state and good qualitative agreement with the first excited state.}
    \label{fig:EnergyTruncations}
\end{figure}

The results obtained using this algorithm are depicted in Figure~\ref{fig:TruncationMap}. As can be inferred from the latter, with $\eta = 1.7 m$ we can simulate the dynamics of the system up to a time $t\approx 3\, t_0$, where $t_0 = 1/\sqrt{m^2+\eta^2}$, with an accuracy of $90 \%$ using a boson truncation of $\Lambda = 3$ ($N=2$). Bearing in mind the limited hardware resources of current quantum computers, we choose this system with a boson truncation of $\Lambda = 3$ (i.e. two bosonic qubits) as our reference system for DQS on an IBM quantum computer. This system shows non-trivial dynamics, thus being a useful testbed for benchmarking purposes.

To examine how the truncation affects the dynamics of the system at hand, we perform classical simulations of the truncated system with $N=1$, $N=2$ and $N=8$ (cf.~Figure \ref{fig:BosonClassicalSim}). Here, the truncation at the large value $N=8$ is chosen in order to ensure that the
system dynamics does not change upon further increase of $N$. As expected, we find a large deviation in the average boson number between the $N=1$ (blue dashed curve) and $N=2$ (orange dotted curve) in the relevant time interval, but a small deviation between the $N=2$ and $N=8$ case (green curve).

The error in the time evolution that stems from an inappropriate boson truncation is caused by the fact that the eigenvalues (energies) and eigenvectors of the untruncated Hamiltonian are poorly approximated. In Figure \ref{fig:EnergyTruncations}, we compare the energies of the ground state $E_0$ (the vacuum state) and the first excited state $E_1$ (the one-boson state) as a function of the interaction strength with different truncations. We find, that the first excited state is poorly approximated in the $N=1$ scenario, which is not surprising, since the truncation should always be larger than the occupation number one wants to simulate. On the other hand, using a truncation of $\Lambda=3$ ($N=2$), we find excellent agreement with the ground-state energy and also reasonably good agreement with that of the first excited state. 

\section{Circuit design}  \label{circ_synth}
In the following we describe the design of quantum circuits emulating the dynamics of the system at hand. 
We start with the constant-depth two-qubit circuit that corresponds to the one-boson exchange case (Section~\ref{twoqubitcirc}),
followed by its three-qubit counterpart in the case of three-boson exchange (Section~\ref{threequbitcirc}). We then 
carry out a detailed analysis of the CNOT-cost estimate for higher boson-number truncations (Section~\ref{CNOTcostHigherTrunc}). Finally, we present the basic aspects of a specific error-mitigation technique -- the (digital) zero-noise extrapolation -- that we will employ in this work (Section~\ref{ZNEbasics}).

Before embarking on the construction of specific circuits, we point out how a specific property of the system
at hand -- namely, the local charge conservation -- allows us to encode the fermion-antifermion sector 
of the problem using only one qubit. Starting from a general expression originating from the JW mapping, namely
\begin{equation}
a^\dagger a + c^\dagger c - 1 = \frac{1}{2}\left(IZ+ZI\right) =
\begin{bmatrix} 
-1 & 0 & 0 & 0\\
0 & 0 & 0 & 0\\
0 & 0 & 0 & 0\\
0 & 0 & 0 & 1
\end{bmatrix}\:,
\end{equation}  
we restrict the system to the subspace with a total charge of $Q=0$. 
Truncating the fermion subspace to states with a vanishing total charge does not affect the behaviour of the system as the Yukawa interaction conserves the total charge, which for our chosen initial state (the vacuum state) is equal to zero.
This allows us to rewrite $a^\dagger a + c^\dagger c - 1$ as
\begin{equation} \label{eqZ}
 a^\dagger a + c^\dagger c - 1 = - Z \:,
\end{equation}
which explains why the fermion-antifermion sector of the problem can be encoded using a single qubit. 

Note that, in theory, one could completely spare the fermion-antifermion sector of the problem as the fermion- and antifermion numbers are seperately conserved in our single-site model (one can see this from Equations~\eqref{ssH0} and \eqref{ssHint}, since both $H_0$ and $H_{\rm int}$ are diagonal in the fermion-number operator). Upon extension to multiple grid points with relativistic momenta, this would no longer hold, as only the charge remains conserved. This is the reason why we do not make use of the particle number conservation at this point and use the fermion-qubit as an additional source of noise.

\subsection{Exchange of up to one boson (two-qubit circuit)}  \label{twoqubitcirc}
The next step is to apply both the fermion and boson mappings from Sections~\ref{EncodeFermions} and \ref{EncodeBosons}, 
respectively, to obtain the qubit Hamiltonian. Here we truncate the bosonic occupation number at $\Lambda = 1$ such 
that $b+b^\dagger = X$. By making use of Equation~\eqref{eqZ}, the two contributions to the total qubit Hamiltonian 
$H_\text{tot} =H_0+H_\text{int}$ are given by
\begin{eqnarray}
H_0 &=& -M\:IZ - \frac{m}{2}\: ZI  + \frac{1}{2}\left(2M + m\right) \:, \label{eq:HQubit1}\\
H_\text{int} &=& -\frac{\eta}{2}\:XZ \label{eq:HQubit2}  \:.
\end{eqnarray}
The spectrum of the free Hamiltonian $H_0$ is symmetrized by dropping the constant term $(2M+m)/2$. 
This changes the time evolution only by an irrelevant 
global phase. 

\begin{figure}[b!]\centering
\includegraphics[width=\textwidth]{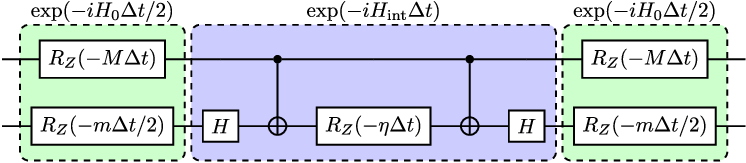}
\caption{\label{fig:CircuitSingleStep}Circuit representing a Trotter step in the 
time evolution of the Hamiltonian $H_\text{tot} =H_0+H_\text{int}$ [cf. Equations~\eqref{eq:HQubit1} and 
\eqref{eq:HQubit2}].}
\end{figure}

The general approach to perform the time evolution of such a Hamiltonian with non-commuting terms 
is based on the Trotter-Suzuki decomposition, also known as Trotterization. To be more specific, 
for the Hamiltonian $H_\text{tot} =H_0+H_\text{int}$ [cf. Equations~\eqref{eq:HQubit1} and \eqref{eq:HQubit2}] 
our starting point is the standard second-order Trotter-Suzuki product formula~\cite{Suzuki:76,Hatano+Suzuki:05}
\begin{equation} \label{2ndOrderTrotter}
e^{-i H_\text{tot} \Delta t} = e^{-i H_0 \Delta t/2} e^{-i H_\text{int} 
\Delta t}e^{-i H_0 \Delta t/2} + O(\Delta t^3)\:,
\end{equation}
where the free Hamiltonian $H_0$ is used for symmetrization as its cost in terms of CNOT gates is zero. 
The circuit for a single Trotter step, based on the ``star'' configuration of CNOT gates 
[see Figure~\ref{fig:ZZZcircuits}(a)], is depicted in Figure~\ref{fig:CircuitSingleStep}.

\begin{figure}[t!]\centering
\includegraphics[width=\textwidth]{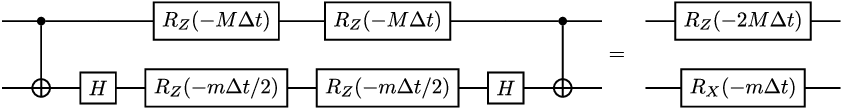}
\caption{\label{fig:CircuitTransZone}Circuit representation of the transition zone between two 
subsequent Trotter steps.}
\end{figure}

In the conventional scenario of using the product formula in Equation~\eqref{2ndOrderTrotter}, one 
obtains a quantum circuit whose depth scales linearly with the number of steps. In other
words, the circuit depth grows with the total simulation time, which in many systems limits feasible 
simulations to relatively short times. We show that for the system at hand -- at least in the 
two-qubit case (i.e.~in the case of up to one boson exchange) -- we can defy this conventional 
scenario and design a constant-depth circuit. To this end, it is instructive to start by analyzing
the transition zone between subsequent Trotter steps, which is represented by the following circuit:

\begin{figure}[b!]
\centering
\includegraphics[width=0.7\linewidth]{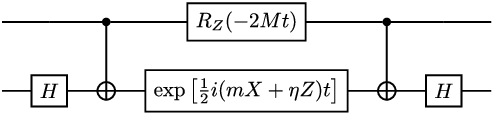}
\caption{\label{fig:ExactCircuitTwoQubits}Exact constant-depth quantum circuit corresponding to the 
evolution governed by the Hamiltonian $H_\text{tot} =H_0+H_\text{int}$ [cf. Equations~\eqref{eq:HQubit1} 
and \eqref{eq:HQubit2}] over the finite time $t$.}
\end{figure}

\begin{figure}[t!]\centering
\includegraphics[width=0.8\linewidth]{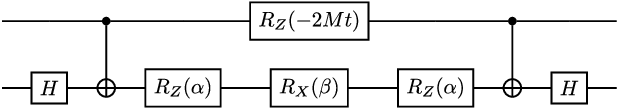}
\caption{\label{fig:TotalTwoQubitCircuit}The full two-qubit circuit emulating the system dynamics in the 
one-boson-exchange case.}
\end{figure}

The simplification of the circuit in Figure~\ref{fig:CircuitTransZone} is made possible by simple
properties of the CNOT gate -- namely, its commutation with a $z$-rotation gate on the control qubit 
and with an $x$-rotation gate on the target qubit, as well as the fact that the CNOT gate is
self-inverse. It is also worthwhile pointing out that the occurrence of $x$-rotation gates on the 
target-qubit wire in the last circuit originates from the identity $R_X(\theta) =H R_Z(\theta) H$, 
which follows from the fact that $X =H Z H$.

From this last circuit, one can easily go back to the exact time evolution by taking the limit $\Delta t \to 0$,
$n\to \infty$ while keeping $n\Delta t\equiv t$ constant. This approach yields the exact constant-depth quantum circuit 
that corresponds to infinitely many small time steps, i.e.~to the finite evolution time $t$; this circuit is shown
in Figure~\ref{fig:ExactCircuitTwoQubits}, where the top wire corresponds to the fermion, the bottom one to the boson. 
It should be emphasized that it is the specific form of the total Hamiltonian $H_\text{tot} =H_0+H_\text{int}$ in 
Equations~\eqref{eq:HQubit1} and \eqref{eq:HQubit2} that allows one to perform circuit compression between different Trotter 
steps such that only two CNOT gates are required to perform an arbitrary number of steps. 

At this point, the problem of decomposing the unitary two-qubit time-evolution operator has been reduced 
to the decomposition of a single-qubit gate. To this end, we recast $\exp\left[i(m X + \eta Z) t/2\right]$ 
by performing a $ZXZ$ Euler decomposition $R_Z(\alpha)R_X(\beta)R_Z(\alpha)$, where
\begin{equation}
 \begin{split}
 \alpha &= -\arctan\left[\frac{\eta}{\omega}\tan\left(\frac{\omega t}{2}\right)\right]\:, \\
 \beta  &= -2\arctan\left[\frac{m}{\sqrt{\eta^2 +\omega^2 \cot^2\left(\frac{\omega t}{2}\right)}}\right]\:, 
\end{split}
\end{equation}
and $\omega^2 = m^2 +\eta^2$. 

The final circuit, which emulates the system dynamics over a finite-evolution time $t$ using only two CNOT gates, then assumes the form depicted in Figure~\ref{fig:TotalTwoQubitCircuit}. In connection with the form of this circuit, it is worthwhile pointing out that circuit compressions of the type utilized here have quite recently been discussed in the context of interacting qubit arrays described by the transverse-field Ising- and $XY$ models~\cite{Kokcu+:22,Peng+:22}, paradigmatic models in condensed-matter physics. These models can be mapped to free fermionic models and are also known to be classically simulatable with polynomial 
resources~\cite{Terhal+DiVincenzo:02}. It is thus interesting that the fermion-boson system under consideration 
offers another, much less common example -- namely, that of an $XZ$-coupled pair of qubits, with mutually unequal 
external fields in the $z$-direction acting on either qubit [cf. Equations~\eqref{eq:HQubit1} and \eqref{eq:HQubit2}] -- where 
such a circuit compression is also possible.

\subsection{Exchange of up to three bosons (three-qubit circuit)}  \label{threequbitcirc}
In order to describe an exchange of up to three bosons, two qubits are required to encode the bosonic Fock space.
Using the bosonic qubit mapping, we modify the Hamiltonian as 
\begin{eqnarray}
H_0 &=& - M\:IIZ - \frac{m}{2}\:ZII - m IZI \:,  \label{eq:ThreeQubitHamiltonian1}\\
H_\text{int} &=& - \frac{\eta}{2} \Big[\frac{1+\sqrt{3}}{2}\:XIZ 
+\frac{1-\sqrt{3}}{2}\:XZZ + \frac{1}{\sqrt{2}}\:(XXZ+YYZ)\Big] \:. \label{eq:ThreeQubitHamiltonian2}
\end{eqnarray}
To perform the time evolution of this Hamiltonian, we start with Trotterization. As this Hamiltonian does not allow 
for the same efficient circuit compression as in the one-boson case, we have to be more careful how to trotterize. 
To reduce the error of a Trotter step to $O(\Delta t^3)$, we once again make use of the second-order Trotter-Suzuki 
product formula [cf. Equation~\eqref{2ndOrderTrotter}], also using the free Hamiltonian $H_0$ for symmetrization.

\begin{figure}[b]
\centering
\includegraphics[width=\textwidth]{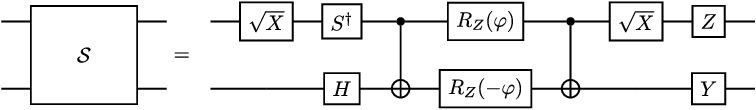}
\caption{\label{fig:CircuitForS} Minimal circuit representation of the unitary $\mathcal{S}$, obtained using the 
Kraus-Cirac decomposition.}
\end{figure}

While a first-order Trotterization of $\exp(-i H_\text{int}\Delta t)$ requires $8$ CNOTs, we will instead construct an exact decomposition with the same gate count. The key idea is that $H_\text{int}$ can be written as a tensor product 
of operators acting locally on the fermionic or bosonic registers.
To calculate the matrix exponential, we may diagonalize these operators separately. The fermionic operator $Z$ is already 
diagonal, leaving us with the task of diagonalizing $b+b^\dagger$. We find $b + b^\dagger =\mathcal S D \mathcal 
S^\dagger$, where in the matrix forms $D$ and $\mathcal S$ are given by
\begin{eqnarray}
D &=& \text{diag}\left(-\lambda_+,\lambda_+,-\lambda_-,\lambda_- \right)\:, \label{exprD}\\
\mathcal S &=& \frac{1}{2}
    \begin{pmatrix}
        -\tilde\lambda_-  & \tilde\lambda_-  & 
        \tilde\lambda_+  & -\tilde\lambda_+  \\
        1 & 1 & -1 & -1 \\
        -\tilde\lambda_+  & \tilde\lambda_+  & 
        -\tilde\lambda_- & \tilde\lambda_-  \\
        1 & 1 & 1 & 1 \\
    \end{pmatrix}\: \label{defineS}\:,
\end{eqnarray}
with $\lambda_{\pm}\equiv\sqrt{3\pm\sqrt{6}}$ and $\tilde\lambda_\pm = \lambda_\pm /\sqrt{3}$. 

\begin{figure}[t]
\centering
\includegraphics[width=0.7\linewidth]{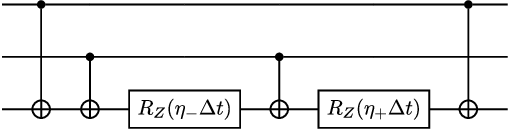}
\caption{\label{fig:CircuitForHint}Quantum circuit representing $\exp(-i H_\text{int, diag} \Delta t)$.}
\end{figure}

The next step is to perform the Kraus-Cirac decomposition of $\mathcal S$, a transformation intimately related 
to the Cartan decomposition of the Lie algebra $su(4)$~\cite{Kraus+Cirac:01,Vatan+Williams:04}. In this manner,
we find that (for a detailed derivation, see Appendix~\ref{KrausCiracDecomp})
\begin{equation}
S = (K_4\otimes K_3)\exp{(i\varphi ZZ/2)}(K_2\otimes K_1) \:, 
\end{equation}
where the matrices $K_1,\ldots,K_4\in\textrm{SU}(2)$ are given by
\begin{eqnarray}
K_1 &=& \frac{e^{i \pi/4}}{\sqrt{2}}\: 
\begin{pmatrix}
i & 1 \\
-i & 1 \\
\end{pmatrix}
= e^{- i \pi/4}S^\dagger \sqrt{X} \:, \nonumber\\
K_2 &=& \frac{e^{i \pi/2}}{\sqrt{2}}\:  
\begin{pmatrix}
            1 & 1 \\
            1 & -1 \\
\end{pmatrix} = \frac{e^{i \pi/2}}{\sqrt{2}}\: H \:,\nonumber\\
K_3 &=& \frac{e^{i \pi/2}}{\sqrt{2}}\:
\begin{pmatrix}
            e^{-i\varphi} & -i e^{i\varphi} \\
            i e^{-i\varphi} & - e^{i \varphi} \\
\end{pmatrix}
= e^{i \pi/2} Z \sqrt{X} R_Z(2\varphi) \:,\nonumber \\
K_4 &=& \begin{pmatrix}
            0 & -1 \\
            1 & 0 \\
\end{pmatrix} = e^{i\pi/2} Y \:,
\end{eqnarray}
with $\varphi=\arctan\left[\sqrt{2}/(1+\sqrt{3})\right]$. The minimal circuit representation 
of $\mathcal{S}$ is depicted in Figure~\ref{fig:CircuitForS}.

The next step is to express $D$ [cf. Equation~\eqref{exprD}] in the Pauli basis:
\begin{equation}
D = \sqrt{\frac{3+\sqrt{3}}{2}}\:ZI + \sqrt{\frac{3-\sqrt{3}}{2}}\:ZZ\:.
\end{equation}
Then, the Hamiltonian $H_\text{int}$ can be recast in the form
\begin{equation}
H_\text{int, diag} = \frac{1}{2} (\eta_- ZZZ + \eta_+ ZIZ)\:,
\end{equation}
where $\eta_{\pm}=\eta\sqrt{(3\pm\sqrt{3})/2}$. The exact circuit for $\exp(-i H_\text{int, diag} \Delta t)$ 
is then given by Figure~\ref{fig:CircuitForHint}.

\begin{figure}[b!]\centering
\includegraphics[width=\textwidth]{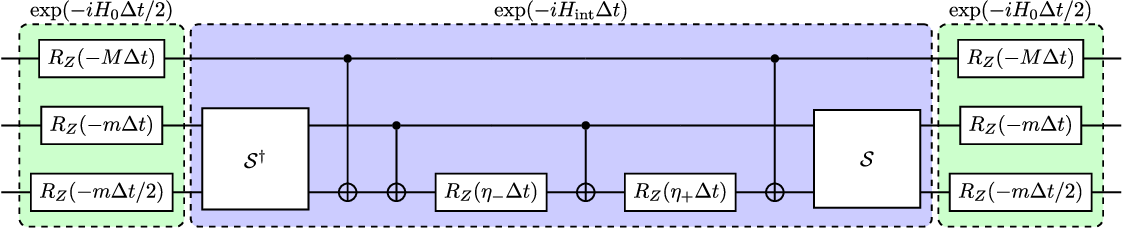}
\caption{\label{fig:ThreeBosonCircuit}Quantum circuit for a single Trotter step in second-order Trotterization 
of the three-qubit Hamiltonian $H_\text{tot} =H_0+H_\text{int}$ [cf. Equations~\eqref{eq:ThreeQubitHamiltonian1} and 
\eqref{eq:ThreeQubitHamiltonian2}]. The bosonic register consists of the mid- and bottom wire.}
\end{figure}

Finally, by combining the circuits shown in Figures~\ref{fig:CircuitForS} and \ref{fig:CircuitForHint} with 
those corresponding to the free Hamiltonian $H_0$, we obtain the circuit that emulates a single Trotter step 
of the system under consideration; this circuit is depicted in Figure~\ref{fig:ThreeBosonCircuit}.

\subsection{CNOT-cost estimation for higher boson-number truncations} \label{CNOTcostHigherTrunc}
For simulations with higher truncation numbers (or more grid points), a more systematic approach 
towards circuit optimization is required due to the rapidly-increasing number of Pauli strings.
Even when considering only a single grid point with a boson-number truncation at $\Lambda=7$ 
(three bosonic qubits), circuit optimization by hand in the problem under consideration 
becomes rather tedious. Namely, for different possible orderings of Pauli strings, we 
obtain different CNOT-gate counts.

\begin{figure}[b!]
\centering
\includegraphics[width=\columnwidth]{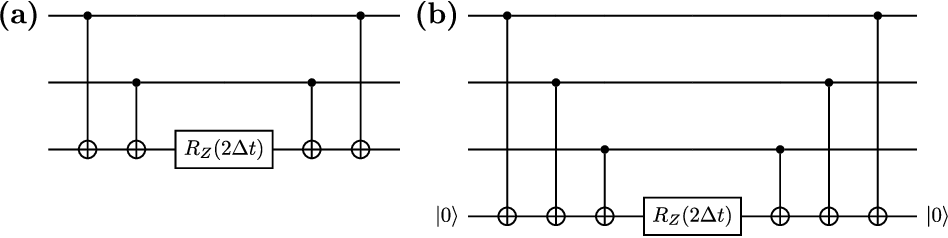}
\caption{\label{fig:ZZZcircuits}Quantum circuits for simulating the Hamiltonian 
$H=ZZZ$ for time $\delta t$ using two different configurations of CNOT gates: (a) star configuration, 
and (b) star+ancilla configuration.}
\end{figure} 

In what follows, we will focus on first-order Trotterization, which means that each Pauli string appears 
exactly once per Trotter step. All considerations in this section are based on the \textit{star + ancilla} layout~\cite{NielsenChuangBook,Gui+:20}, which is depicted in Figure~\ref{fig:ZZZcircuits}(b). Using this 
circuit layout, we can simulate the time evolution of any Hamiltonian of the form
\begin{equation}
H = \bigotimes_{k=0}^{n-1}\sigma_{c(k)}^k \:,
\end{equation}
where $\sigma_{c(k)}^k\in \{I, X, Y, Z\}$ is a Pauli operator acting on the $k$-th qubit and 
$H$ is given by a single Pauli string. The design of a quantum circuit emulating the time evolution 
governed by such a Hamiltonian is explained in Appendix~\ref{DistanceMetric}. For Hamiltonians with multiple 
non-commuting Pauli strings, we append the same circuit structure for each string and justify the circuit 
decomposition through the Trotter product formula.  

Qubits with identity operations can be disregarded and qubits with $X$ or $Y$ are transformed 
to $Z$ using $X = HZH$ and $Y = SHZHS^\dagger$. Whenever we deal with multiple non-commuting 
Pauli strings, we use Trotterization to recover the circuit structure from Figure~\ref{fig:ZZZcircuits}(b)
for each string.   

For the sake of simplifying the analysis of 
the CNOT-cost in this section, we ignore the fermionic qubit in the interaction Hamiltonian $H_\text{int}$, and focus entirely on the bosonic term $H_\text{b}
\propto b+b^\dagger$. In other words, our goal is therefore to estimate the CNOT-cost of the circuit representation 
of $\exp\left(-i H_\text{b}\Delta t \right)$ for an arbitrary truncation $\Lambda$. We restrict 
ourselves to the case where the truncation can be expressed as $\Lambda = 2^N - 1$, where $N$ is 
an integer corresponding to the amount of qubits forming the bosonic register. Let $S_N = \{P_1, 
P_2, \dots, P_k\}$ denote the set of Pauli strings required to encode $b+b^\dagger$. In Appendix~\ref{GeneratePauliStrings}, we show that the number of Pauli strings $k$ for a truncation to $N$ 
qubits is given by
\begin{equation}
k = |S_N| = N 2^{N-1} \:.
\end{equation}

\begin{figure}[t]
\centering
\includegraphics[width=\textwidth]{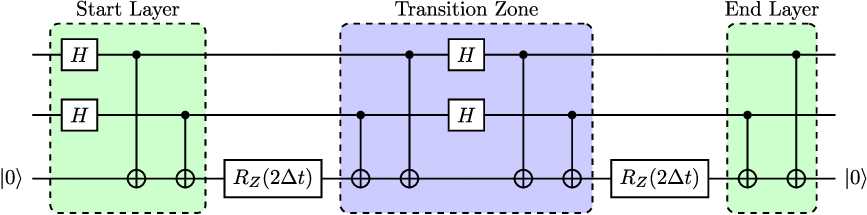}
\caption{\label{fig:TransitionZoneGraphic}Time evolution circuit for the Hamiltonian $H=XX+ZZ$ 
using the Star+Ancilla layout. We define the zones on the left (right) of the first (last) 
$R_Z$-gate as the start (end) layer. A zone between two subsequent $R_Z$-gates is referred to 
as a transition zone.}
\end{figure}

To find the optimal order of these strings, we use an approach based on an analogy with the TSP, adapting a technique proposed in Ref.~\cite{Gui+:20}. The formulation of this optimization task as a TSP works as follows. We first define a weighted, fully-connected graph $G$ with $k$ nodes, where each node represents a Pauli string. We then set the edges between all nodes $i$ and $j$ with a weight given by $|P_i-P_j|_\text{CNOT}$, where $|P_i-P_j|_\text{CNOT}$ denotes the CNOT-cost to implement the transition between the circuits representing $\exp\left(-i P_i\Delta t \right)$ and $\exp\left(-i P_j\Delta t \right)$. A graphical illustration of the definition of the transition zone is provided in Figure~\ref{fig:TransitionZoneGraphic}. 
In Appendix~\ref{DistanceMetric} we prove that the CNOT-cost to implement such a transition is precisely given by the Hamming distance\begin{equation}\label{eq:CNOTDistMod}
|P_1-P_2|_\text{CNOT} := \sum_{i \in [N]} 1_{P_1[i]\neq P_2[i]} = |P_1-P_2|_\text{Ham}\:.
\end{equation}
Note that this distance metric differs by the one used in Ref.~\cite{Gui+:20}, since we take additional gate identities into account, which can significantly reduce the CNOT-cost, as explained in Appendix~\ref{DistanceMetric}.

\begin{figure}[b]
\centering
\includegraphics[width=0.45\columnwidth]{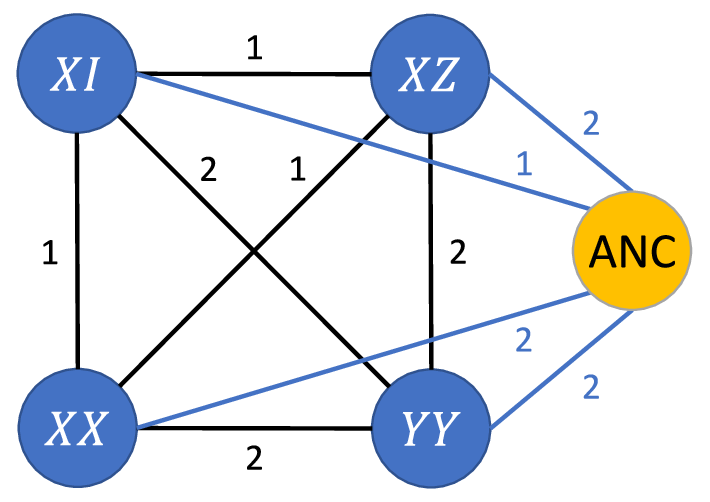}
\caption{\label{fig:GraphLambda3}Graph representation of the set of Pauli strings $\{XI, XZ, XX, YY\}$. 
The weights between two Pauli string nodes (black edges) correspond to the Hamming distance (c.f.~Equation~\eqref{eq:CNOTDistMod}), 
while the weight between a Pauli string node and the ancilla (ANC) node (blue edges) is the Hamming weight (c.f.~Equation~\eqref{eq:hamming_weight})
}
\end{figure}

TSP aims to minimize the cost of a closed path visiting all nodes exactly once. For a circuit made of $k$ Pauli strings, there are $k-1$ transition zones. A closed path on the graph described above would however consist of $k$ transition zones. In Ref.~\cite{Gui+:20}, this issue is fixed by inserting an ancilla node which is connected to all nodes with zero weights. Note that this approach ignores the CNOT-cost of the start- and end-layers of the Trotterization circuit. In order to take also this cost into account, we find that the number of CNOTs to implement these layers is given by the Hamming weight
\begin{equation}\label{eq:hamming_weight}
|P_j|_\text{Ham} = \sum_{i \in [N]} 1_{P_j[i]\neq I}\:.
\end{equation}
Therefore, we use the above equation to calculate the weights for the edges connecting the nodes of the original graph with the ancilla node.

The closed path on our modified graph now contains $k+1$ edges, of which two correspond to the start- 
and end layers, meaning that we have $k-1$ transition zones, as desired. 

Finally, to build the graph for the problem at hand, 
we need to know the Pauli strings explicitly. We generate the sets of Pauli strings recursively using 
\begin{eqnarray}
S_{N+1} &=&  S_N \otimes I \cup S_N \otimes Z \cup S(\sigma_+^{\otimes N}\otimes\sigma_-  + \sigma_-^{\otimes N}\otimes \sigma_+) \:,\label{eq:PauliStringsRec}
\end{eqnarray}
with $S_1 = \{X\}$, where the tensor product is performed element-wise (e.g., $S_N \otimes Z = 
\{P_1 \otimes Z,\ldots, P_k \otimes Z\}$) and $S(O)$ denotes the set of Pauli strings building 
the operator $O$. A naive expansion of the last set $S(\sigma_+^{\otimes N}\otimes\sigma_-  + 
\sigma_-^{\otimes N}\otimes \sigma_+)$ gives us $2^{N+1}$ different strings where each character 
can be either $X$ or $Y$. But as the operator is Hermitian, all strings with imaginary coefficients 
cancel out. These strings are precisely the ones with an odd number of $Y$s. This leaves us with a 
subset containing $2^N$ strings with an even number of $Y$s. An example of the completed graph for 
a truncation with two qubits is shown in Figure~\ref{fig:GraphLambda3}.

\begin{figure}
\centering
\includegraphics[width=0.7\columnwidth]{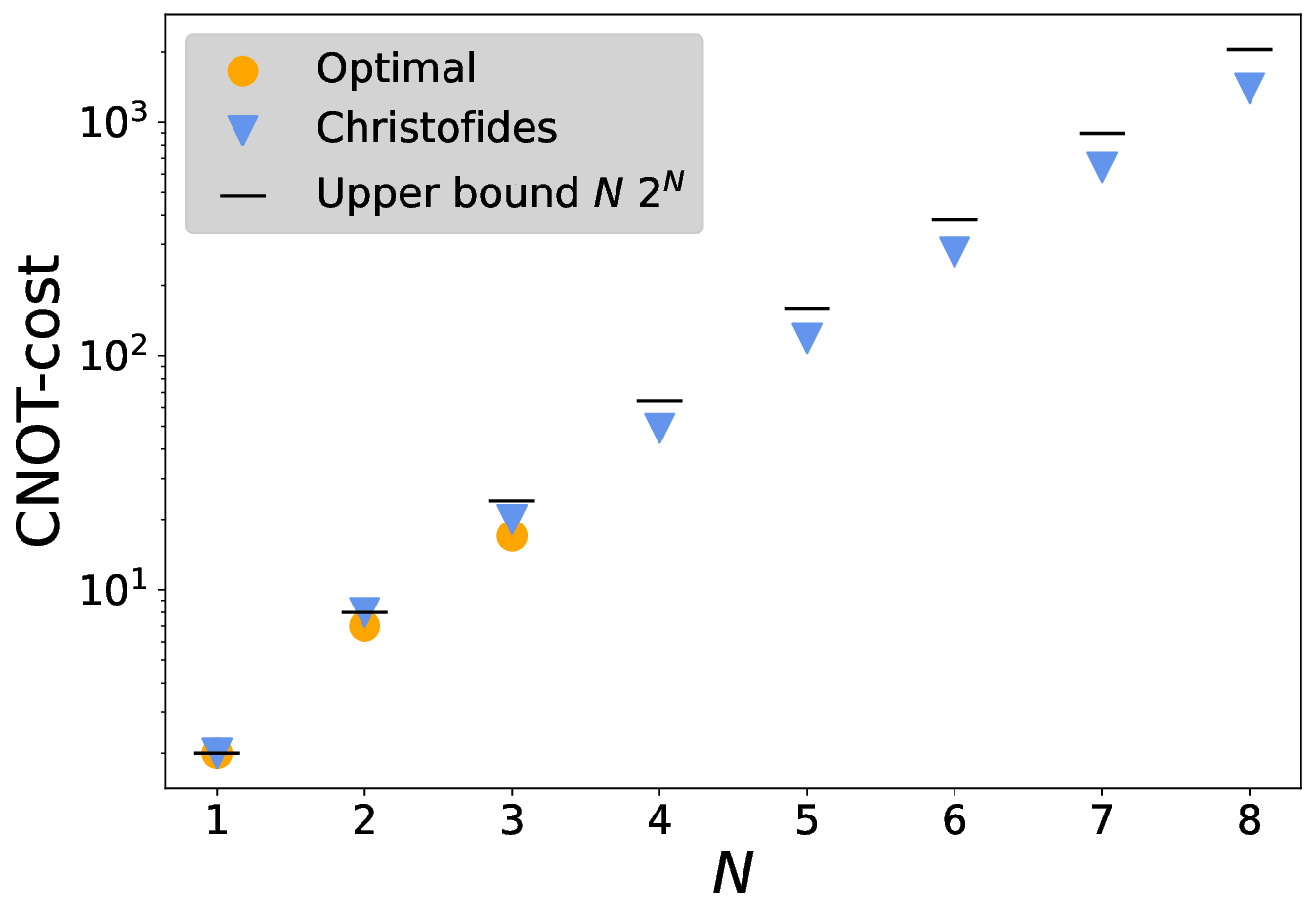}
\caption{\label{fig:CNOTcostThreeMeth}CNOT-cost for implementing the bosonic time evolution as a function of the qubit number $N$. Our algorithm achieves significant better results than the upper bound of $N 2^{N}$.}
\end{figure}

To solve the TSP exactly, we make use of the Bellman-Held-Karp dynamic-programming algorithm~\cite{Bellman:62,Held+Karp:62},
which scales in time with $\mathcal{O}(k^22^k)$. Due to the exponential growth of the number of Pauli strings, we 
are limited to truncations with $N\leq 3$ ($\Lambda \leq 7$). Because we are dealing with a metric graph, TSP can be 
$1.5$-approximated in polynomial time $\mathcal{O}(k^3)$ using the Christofides algorithm~\cite{Christofides:76}. 
We utilize this heuristic to approximate the CNOT-cost for truncations with $N\leq 8$ ($\Lambda \leq 255$). 
Finally, we compare the obtained results to an upper bound on the CNOT-cost derived in Appendix~\ref{CNOTcostBound}. 
The results obtained using all three methods are illustrated in Figure~\ref{fig:CNOTcostThreeMeth}. While, needless 
to say, the Bellman-Held-Karp algorithm yields the best results, the Christofides heuristic outperforms the upper 
bound we propose. This suggests that a tighter bound on the CNOT-cost can be found.

\subsection{Error mitigation by digital zero-noise extrapolation} \label{ZNEbasics}
Current NISQ devices, due to the limited number of qubits, do not allow one to perform 
full-fledged quantum error correction.
As an alternative to error correction, various error-mitigation strategies have been proposed~\cite{Kern+:05,
Kern+Alber:05,Temme+:17,Li+Benjamin:17,
McArdle+:19,Czarnik+:21,Guo+Yang:22}. They all aim to identify the noiseless signal from a large number of noisy experimental repetitions. 

In the present work, we focus on one specific error-mitigation strategy, namely digital {\it zero-noise extrapolation}~(ZNE)~\cite{Temme+:17,Li+Benjamin:17,Giurgica-Tiron+:20}. Much like other error-extrapolation methods, ZNE make use of expectation values $E(\lambda)$, obtained at multiple different physical noise levels that are quantified by the dimensionless scale factor $\lambda$
(where $\lambda=1$ is the actual noise level of the physical hardware), to infer the noiseless expectation value $E(\lambda=0)$. 

The basic idea of ZNE is to scale the noise of a quantum computation by means of {\em unitary folding}, i.e., replacing a unitary circuit $U$ by
\begin{equation}
 U \rightarrow U (U^{\dagger}U)^n \:,
\end{equation}
where $n$ is a positive integer. While in an ideal circuit -- where $U^{\dagger}U =\mathbbm{1}_D$, with $D$ being the dimension of the Hilbert space associated to all the qubits of the circuit -- this folding operation has a trivial effect, one can expect that on a realistic (noisy) quantum computer the same operation amounts to increasing the noise. 

The unitary folding can be implemented in the guise of {\em circuit 
folding} and {\em gate folding}.
In this work, we use circuit folding. 
To illustrate this approach, let us assume that we have a quantum circuit emulating the time evolution of a system, where each Trotter step (corresponding to the time $\delta t$) is described by the unitary $U$. 
Then, e.g., the state of the system at time $t=3\delta t$ is given by $\ket{\psi(t)} = U U U \ket{\psi(t=0)}$. 
To increase the noise in the circuit in order to perform ZNE we are replacing $U U U$ by $U U U (U^{\dagger}U^{\dagger}U^{\dagger}) U U U$ (corresponding to $\lambda = 3$) or by 
$UUU((U^{\dagger}U^{\dagger}U^{\dagger})UUU)^2$ (corresponding to $\lambda = 5$). 

Assuming that the noise model for the relevant quantum circuit is that of a global depolarizing channel~\cite{NielsenChuangBook}, the density 
matrix of the system can be represented as 
\begin{equation}\label{eq:polarization_noise}
 \rho \xrightarrow{\textrm{noisy circuit}} p\: U\rho U^{\dagger}
 +(1-p)\mathbbm{1}_D/D \:,
\end{equation}
where $p \equiv \prod_{j}p_j$ is the product of the gate-dependent noise parameters $p_j\in [0,1]$ for all the gates in the given circuit. If all the input gates are folded exactly $n$ times (the corresponding value of the scale factor is $\lambda = 2n+1$), in both the circuit- and gate-folding cases one obtains 
\begin{equation}
 \rho \xrightarrow{\textrm{noisy + unitary folding}} p^{\lambda}\: U\rho U^{\dagger}
 +(1-p^{\lambda})\mathbbm{1}_D/D \:,
\end{equation}
i.e., an exponential scaling -- in terms of the noise level $\lambda$ -- of the depolarizing parameters $p_j$ for each gate ($p_j \rightarrow p_j^{\lambda}$). Therefore, under the aforementioned initial assumption of depolarizing channel (which commutes with unitary operations) the unitary folding amounts to an exponential dependence of any expectation value $E(\lambda)$:
\begin{equation}
 E(\lambda) = a + b p^{\lambda} \:.
\end{equation}
The coefficients $a$ and $b$ can be 
determined by fitting the obtained numerical
results to the last exponential ansatz, which
is followed by a straightforward extrapolation 
to the noiseless case ($\lambda=0$).

In this work, we use a more general polynomial extrapolation to find the noiseless limit. 
In this case one uses a polynomial of degree $d$ as ansatz:
\begin{equation}
 E(\lambda) = E_0 + \sum_{n=1}^d c_n \lambda^n,
\end{equation}
such that the intercept $E_0$ corresponds to the zero-noise limit.
In the following, we consider extrapolations with $d=2$ and $d=3$.

\section{DQS on IBM Q: Results and Discussion}  \label{ResDisc}
In what follows, we present and discuss the results 
obtained by running the designed quantum circuits 
emulating the system dynamics -- both for the cases 
of up to one and three boson-exchange processes -- on 
IBM Q. We start with the results obtained for 
the Loschmidt echo (Section~\ref{LoschmidtRes}). We then present the results obtained for boson occupation numbers (Section~\ref{BosonOccupation}) and adiabatic 
state preparation (Section~\ref{sec:AQC}).

We benchmark the accuracy of our DQS by comparing the results obtained on the IBM Q processor \cite{ibmq} \texttt{ibmq\textunderscore manila} with 
those resulting from a numerically-exact treatment of the quantum dynamics governed by the truncated Hamiltonians of the model under consideration on a classical computer.

To demonstrate that our simplified, single-site model of scalar Yukawa coupling exhibits nontrivial quantum dynamics, we perform 
DQS of this model taking the vacuum state as the initial state of the boson-fermion system at hand. 
We construct the states $|\psi(t)\rangle$ of the system at different times $t$ by applying the designed two- and three-qubit circuits that emulate the system dynamics following a Yukawa-interaction quench at $t=0$. Finally, 
for each of those constructed states we perform measurements that give us access to the desired quantities, bearing 
in mind the relevant expressions in the Pauli basis [cf. Equations~\eqref{eq:JWMapping} and \eqref{eq:BosonPartNumber}]. 

\begin{figure}
\centering
\includegraphics[width=0.45\columnwidth]{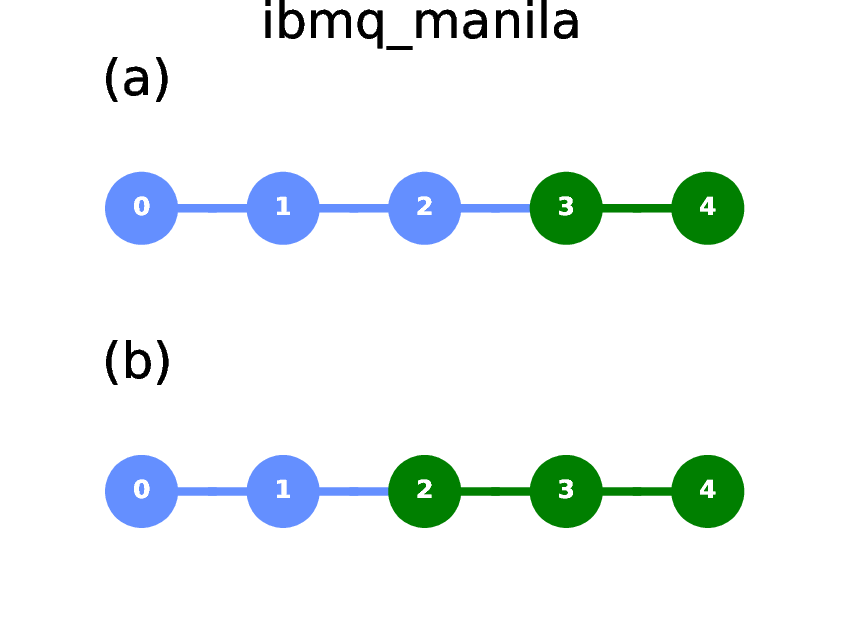}
 \caption{\label{fig:CouplingMaps2}
Qubit-connectivity graphs (coupling maps) of 
\texttt{ibmq\textunderscore manila} for (a) two-qubit case (with qubits 3 and 4 used in the actual implementation), and (b) three-qubit case (with qubits 2, 3, and 4 used in the implementation). 
}
\end{figure}

Given that the DQS of the single-site boson-fermion model under consideration only requires two- and three-qubit systems [cf. Section~\ref{circ_synth}], for benchmarking purposes it was sufficient to make use of the freely available, five-qubit processor \texttt{ibmq\textunderscore manila}~\cite{ibmq}. The qubit-connectivity graph (coupling map) of this quantum processor, with the corresponding enumeration of physical qubits, is depicted in Figure~\ref{fig:CouplingMaps2}. 

Generally speaking, the errors in numerical experiments on IBM Q devices strongly depend on the number of CNOT 
gates in the corresponding quantum circuit. The principal reason for this is that the CNOT-gate error is an order 
of magnitude larger than that of single-qubit gates. Another reason is that the CNOT-gate time is much longer
than that of its single-qubit counterparts, which leads to the accumulation of errors due to energy relaxation 
and dephasing, this two processes being quantified by the respective decoherence times $T_1$ and $T_2$~\cite{Wendin:17}. 
Therefore, in order to minimize the error, in our DQS we make use of the connected subsets of qubits with 
the smallest average CNOT error according to the calibration data from Ref.~\cite{ibmq} from which we found that for our three-qubit circuits it is beneficial to use the qubit pairs $(2,3)$ and $(3,4)$.

The simulation parameters used for the following benchmarks are $M/m=7$ (nucleon/pion mass ratio), $\eta/m = 1.7$ and $t_0\equiv 1/\sqrt{m^2 + \eta^2}$. For the calculations of all the following observables, we use the \textit{Estimator primitive} as well as the \textit{Sampler primitive} provided in \texttt{qiskit\textunderscore runtime} \cite{QiskitRuntime}. This allows to use post-correction of all observables. We set $\texttt{resilience\textunderscore level}=1$, which performs a readout error correction on the measured observable.
We ran each circuit on the IBM Q processor with 8192 shots.

The code used to produce the plots in this section can be found in a supplementary file of this paper.

\subsection{Loschmidt echo and state fidelity}
\label{LoschmidtRes}
We start by evaluating the Loschmidt echo, which in the problem under consideration is given by 
[cf. Equation~\eqref{LoschmidtAmp}]
\begin{equation}
\mathcal{L}(t)=|\langle\psi_{t=0}|\:e^{-i (H_0 + H_{\textrm{int}})t}
\:|\psi_{t=0}\rangle|^{2} \:,
\end{equation}
and can be calculated performing the reconstruction of the state of the system at time $t$ using quantum-state-tomography 
algorithms~\cite{Smolin+:12}. In the problem at hand, the time-evolution (DQS) circuits [cf. 
Section~\ref{circ_synth}] -- which emulate the dynamics governed by the total Hamiltonian $H_{\textrm{tot}}
=H_0 + H_{\textrm{int}}$ of the system after a Yukawa-interaction quench at $t=0$, starting from the initial 
state $|\psi_{t=0}\rangle\equiv|\psi_0\rangle$ -- play the role of the required state-preparation circuits. 

Generally speaking, the objective of quantum-state tomography is to reconstruct the density matrix $\rho$ that 
corresponds to the actual state of the system~\cite{NielsenChuangBook, Gross+:10, Gross:11}. Because our quantum circuits do not involve any measurements, 
the state of the system can be described using the state-vector formalism (recall that the density matrix corresponding
to the pure state $\ket{\psi}$ is given by $\rho = \ket{\psi}\bra{\psi}$). In general, to carry out quantum-state 
tomography on an $N$-qubit system one has to perform $3^N$ measurements for a generic mixed state~\cite{Altepeter+:04,Smith+:21}. 

Throughout this work, we used the pre-implemented tomography method from the \texttt{qiskit-experiments} package \cite{QiskitExperiments}, 
which comes with different tomography fitters. 
More specifically, we used the \textit{linear\_inversion} fitter, based on the work of Ref.~\cite{Smolin+:12}, which is an efficient method for computing the maximum-likelihood quantum state given a set of measurement outcomes in a complete orthonormal operator basis.

Given the reconstructed density matrix of the final state, as well as the density matrices of the initial state
and the ideal final state (obtained through classical simulation), we compute the Loschmidt echo and the fidelity 
of the evolved state starting from the general expression for the quantum-state fidelity~\cite{Uhlmann:76} 
\begin{equation}
\mathcal{F}(\rho, \sigma) = \left(\text{Tr}
(\sqrt{\sqrt{\rho}\:\sigma\sqrt{\rho}})\right)^2 \:,
\end{equation}
where $\rho$ and $\sigma$ are the density matrices of the two mixed quantum states. In the special case of pure 
states -- of interest in the problem at hand -- the last general expression reduces to $\mathcal{F}(\rho,\sigma) = 
|\braket{\psi_\rho|\psi_\sigma}|^2$ (here $\rho\equiv \ket{\psi_\rho}\bra{\psi_\rho}$ and $\sigma\equiv 
\ket{\psi_\sigma}\bra{\psi_\sigma}$ are the density matrices corresponding to the pure states $\ket{\psi_\rho}$ 
and $\ket{\psi_\sigma}$, respectively).
\begin{figure}
    \centering
    \includegraphics[width=0.45\columnwidth]{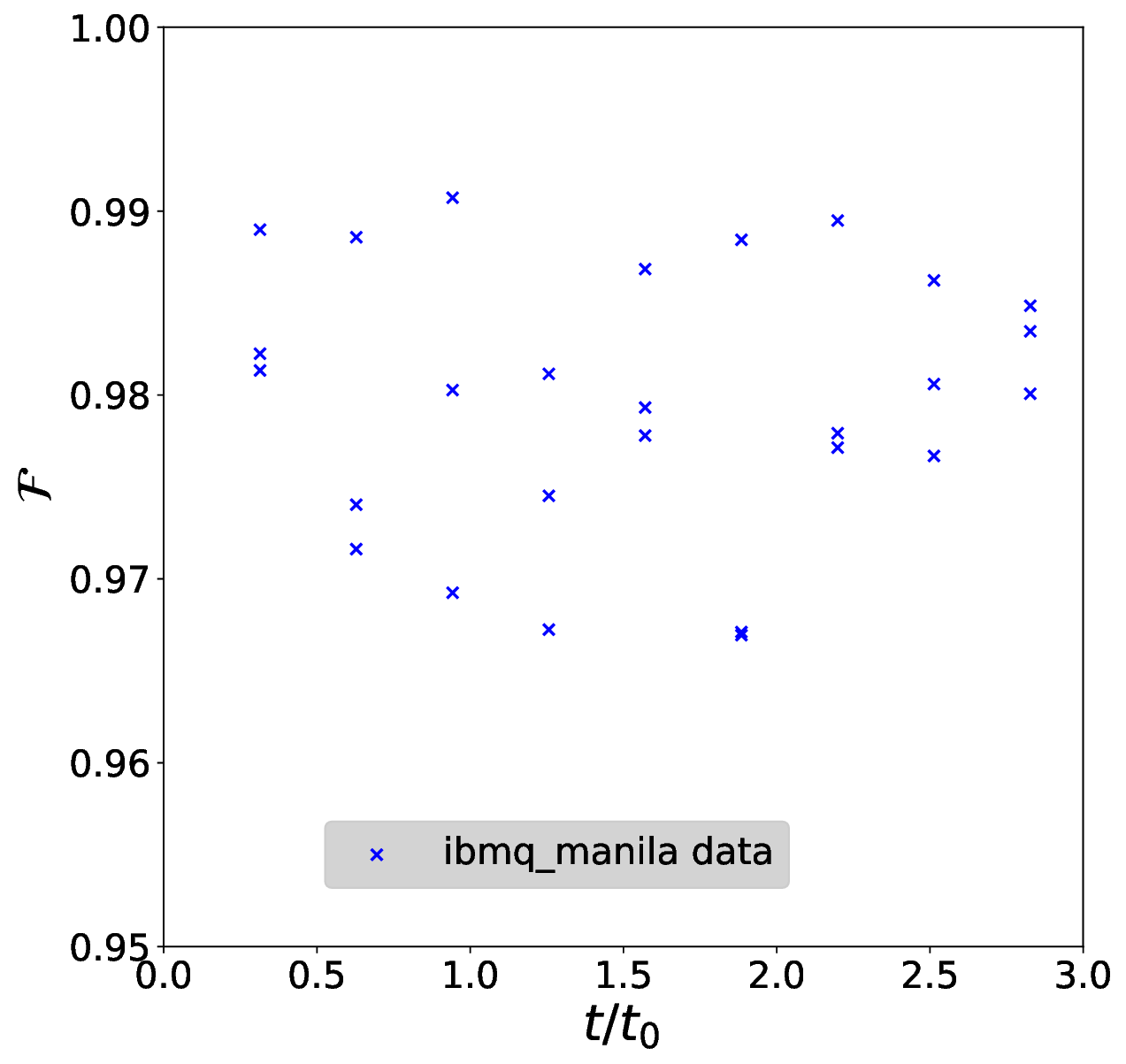}
    \includegraphics[width=0.45\columnwidth]{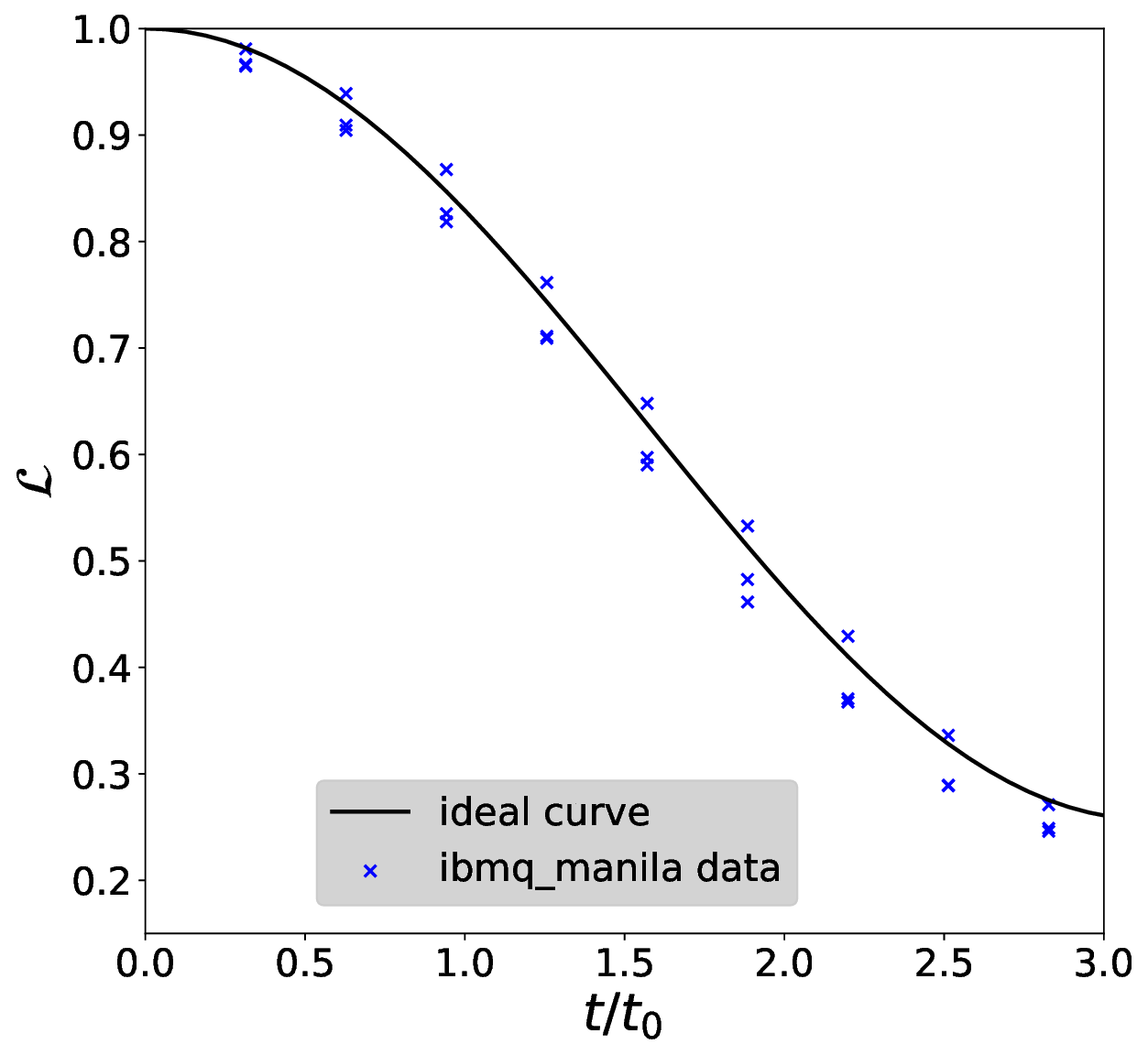}
    \caption{State fidelity $\cal F$ (left panel) and Loschmidt echo $\cal L$ (right panel) as function of the dimensionless time $t/t_0$ for the two-qubit circuits. The fidelity remains constantly high during the whole simulated time interval, which is a consequence of the constant depth circuit construction. The measured Loschmidt echo shows excellent agreement with the theory curve. Each experiment was repeated three times.}
    \label{fig:2q_fidelity}
\end{figure}

\begin{figure}[t!]
    \centering
    \includegraphics[width=0.55\columnwidth]{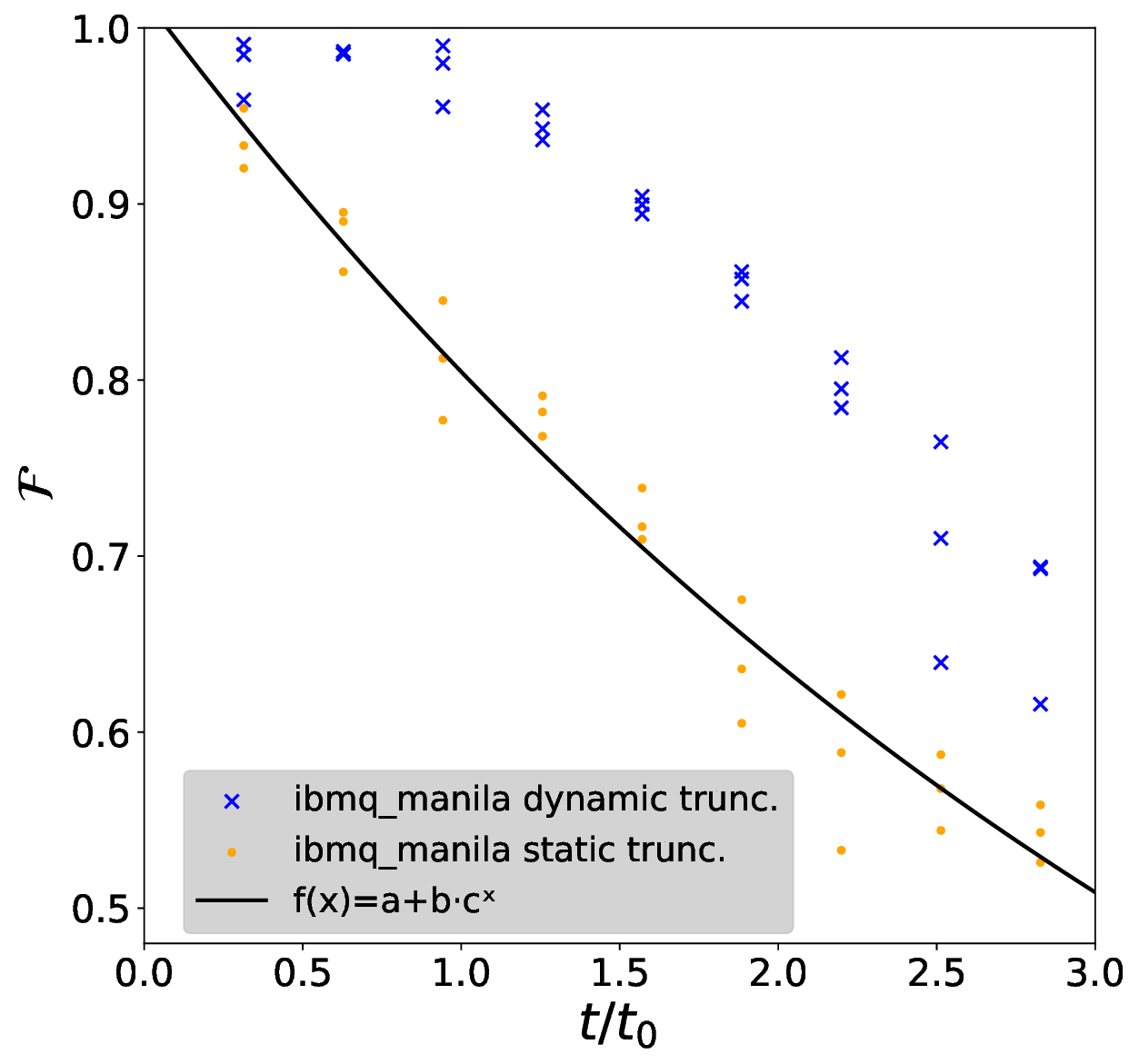}
    \caption{State fidelity $\cal F$ as a function of the dimensionless time $t/t_0$ comparing the dynamic with the static truncation. As can be seen, the fidelity for the dynamic truncation circuits stays on a plateau for the first three steps, for which we used the two qubit circuits, and then starts to fall. The decrease of the fidelity can be explained by the increase of noise in the circuit. We fit this decrease using a exponential model, from which we can deduce the average CNOT gate errors of \texttt{ibmq\textunderscore manila}.}
    \label{fig:3q_fidelity}
\end{figure}

In the following we make use of two different simulation schemes in order to increase the truncation number in the time evolution of the system. In the first one (\textit{static truncation}), we use the same boson-truncation throughout the entire time evolution process. In the second one (\textit{dynamic truncation}) we start the time evolution with a lower truncation number (here $N=1$) and then increase after a critical time, according to Figure~\ref{fig:TruncationMap}. While, in principle, this increases the overall truncation error, the effect is negligible compared to the benefit of simulating time steps with lower-depth circuits. 

In our \textit{dynamic truncation} simulation scheme, we use the two-qubit circuit for the first three time steps, and append the Trotterized three-qubit circuit starting from the fourth time step. We increase the truncation earlier than indicated in Figure~\ref{fig:TruncationMap}, since we want to reach a state fidelity much higher than 90\% at the point where the truncation is increased, such that the target accuracy of 90\% can be achieved in the end. 

The obtained results for the state fidelities are shown in Figures~\ref{fig:2q_fidelity} (left panel) and \ref{fig:3q_fidelity} for the two- and three-qubit circuits, respectively. We repeat each experiment three times. As expected, for the two-qubit circuits the fidelity remains constant at roughly $0.98$ on average for all ten time steps. This is the effect of our constant-depth circuit (i.e., circuit compression), leading to a high fidelity at all times. For the 
three-qubit case we observe a fast decrease of the state fidelity from the beginning in the static truncation simulation, while for the dynamic truncation this decrease starts after $3$ steps only, when we have to use the three-qubit circuits to simulate the system. This decrease of the fidelity can be attributed to the increase of noise in the quantum circuit when more gates are added. Under the assumption of a depolarization model [cf. Equation~\eqref{eq:polarization_noise}], the fidelity drop can be fit with an exponential function,
\begin{equation}
    f(n) = a+b\cdot c^n \: ,
\end{equation}
where $n$ is the number of time steps. From the value of $c$ one can estimate the average error of applying the unitary evolution for one time step, which is given by $1-c$. From our fit, we find
\begin{equation}\label{eq:cval}
    c = 0.925\pm 0.025 \:.
\end{equation}

\begin{figure}[b!]
    \centering
    \includegraphics[width=0.95\columnwidth]{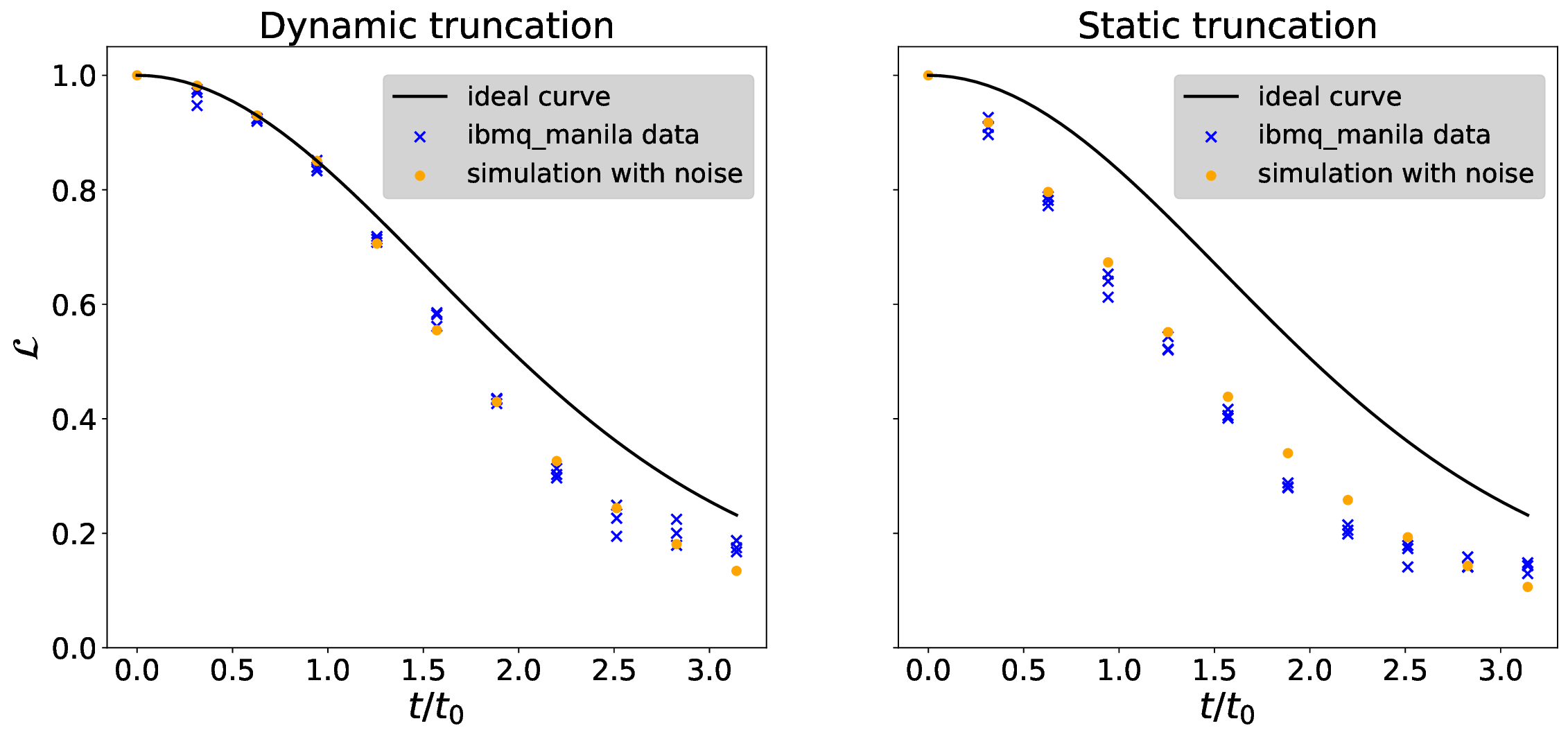}
    \caption{Loschmidt echo $\mathcal{L}$ as a function of the dimensionless time $t/t_0$ for the dynamic truncation time-evolution (left panel) and the static truncation time-evolution (right panel). The data from the quantum computer (blue crosses) is shifted compared to the ideal theory curve. This shift can be explained by polarization noise, which we modeled according to Equation~\eqref{eq:polarization_noise} (orange dots).}
    \label{fig:3q_loschmid}
\end{figure}

In the right panel of Figure~\ref{fig:2q_fidelity} and Figure~\ref{fig:3q_loschmid} we show the results for the Loschmidt echo for the two- and three-qubit circuits. In the two-qubit case we find a good agreement between the ideal theory curve and the experimental data from the quantum processor. For the three-qubit case we find a deviation, which increases with the number of time steps, as expected due to the accumulation of noise. Using the fit-parameter from the fidelity experiments, we try to model the effect of depolarization noise on the Loschmidt echo (orange points in Figure~\ref{fig:3q_loschmid}). We do this by calculating the noisy Loschmidt echo with the depolarized density matrix
\begin{equation}
\rho \rightarrow c^n\rho + \frac{(1-c)^n}{8}
\mathbbm{1}_{8},
\end{equation}
where $c$ is given by Equation~\eqref{eq:cval} and $n$ denotes the $n$-th step. As can be seen from that plot, this model can qualitatively describe the deviation from the data compared to the ideal theory curve in the dynamic truncation time evolution as well as in the static one.

\subsection{Boson occupation numbers\label{BosonOccupation}}
\begin{figure}[b!]
    \centering
    \includegraphics[width=0.6\columnwidth]{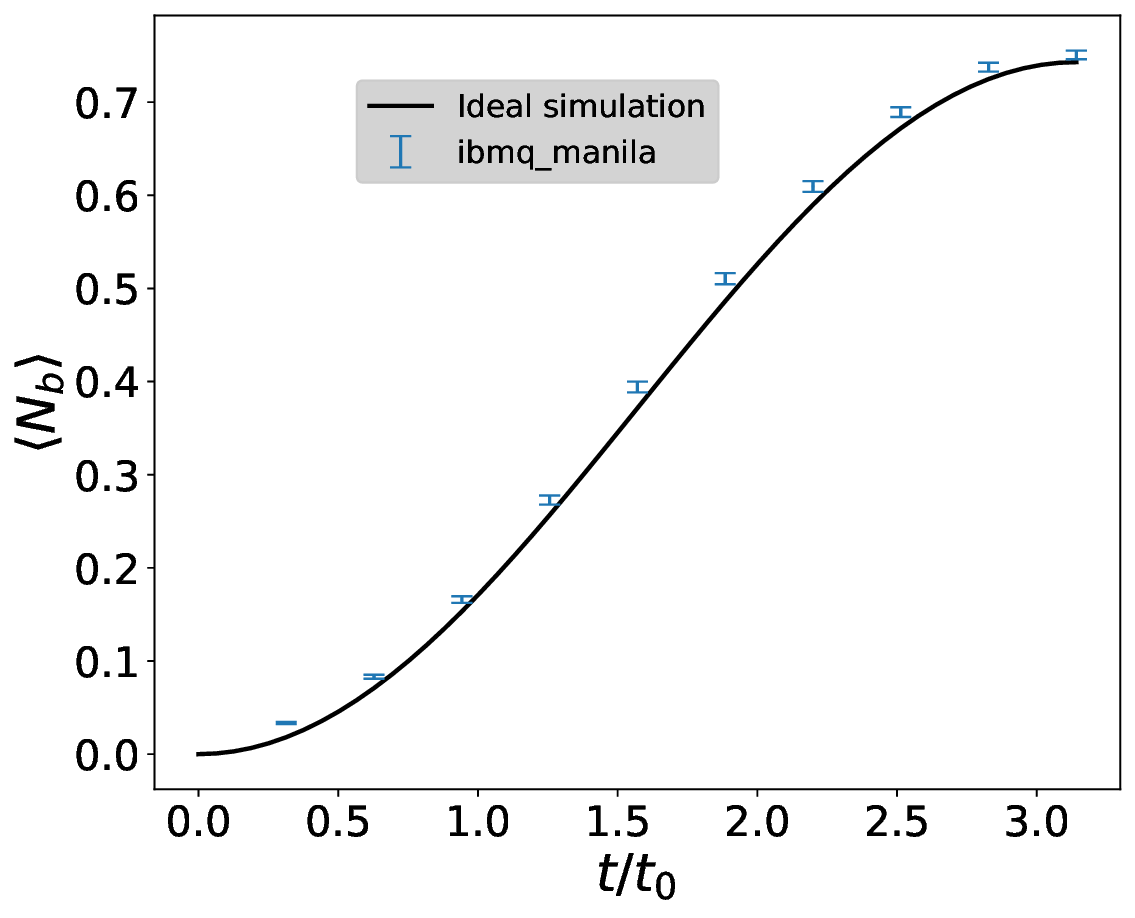}
    \caption{Expected boson number $\braket{N_b}$ as a function of the dimensionless time $t/t_0$ using one bosonic qubit. The error bars correspond to shot noise assuming a Gaussian distribution.}
    \label{fig:2Q_Occupation}
\end{figure}

The obtained results in the two-qubit case are illustrated in Figure~\ref{fig:2Q_Occupation}, which shows the expectation values of the boson-number operator at an arbitrary time $t$ after the quench. Apparently, the compressed circuit with only two CNOT gates
[cf. Section~\ref{twoqubitcirc}] shows good agreement with the exact dynamics. The error-bars in Figure~\ref{fig:2Q_Occupation} show the effect of shot noise in the quantum computation under the assumption of a normal 
distribution.

\begin{figure}[t!]
    \centering
    \includegraphics[width=1.0\columnwidth]{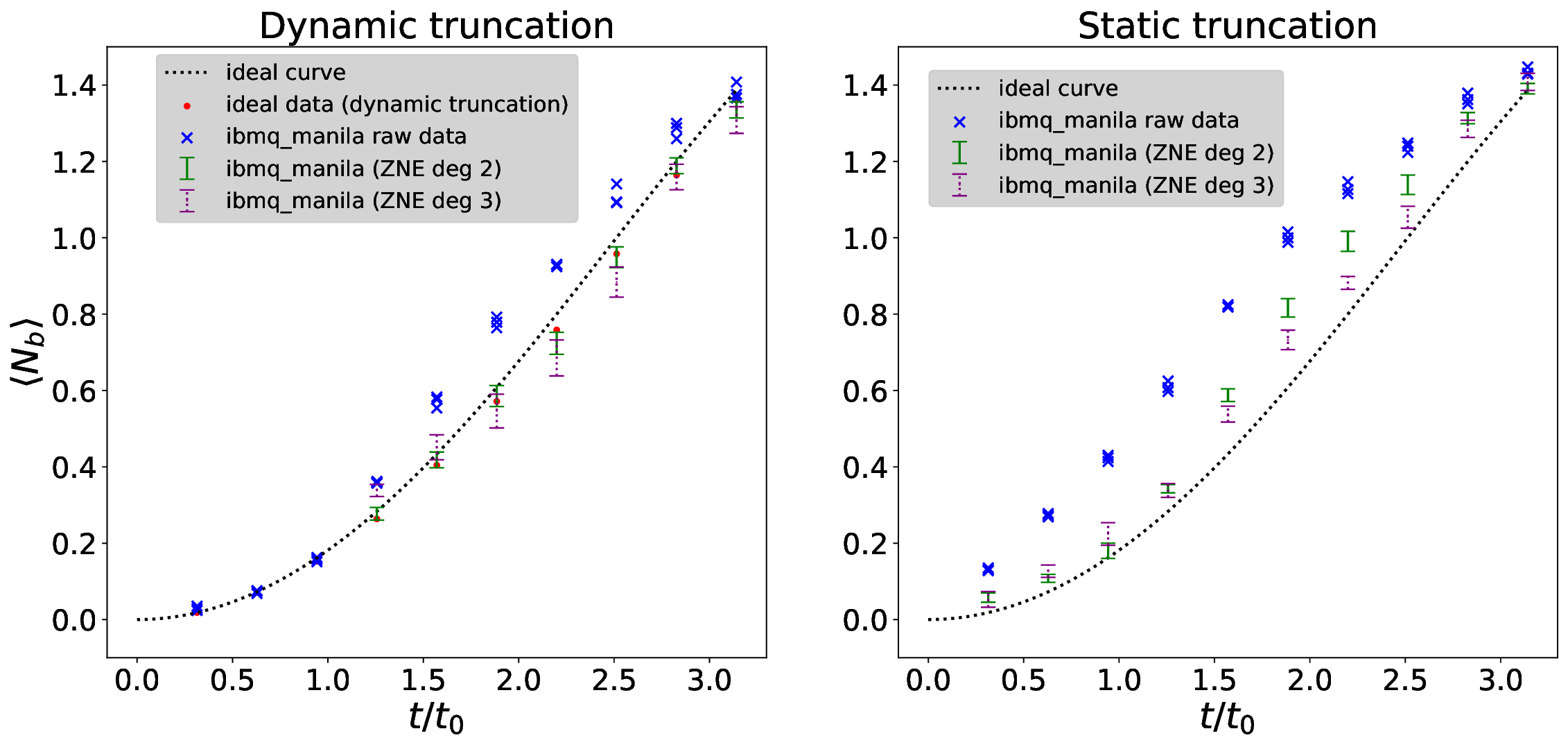}
    \caption{Expected boson number $\braket{N_b}$ as a function of the dimensionless time $t/t_0$ for dynamic truncation (left panel) and static truncation (right panel) circuits. While for the raw data the dynamic truncation scheme gives much better results than the static truncation, the ZNE is able to reduce the error in all cases. \label{fig:3Q_Occupation}}
\end{figure}

\begin{figure}
    \centering
    \includegraphics[width=1.0\columnwidth]{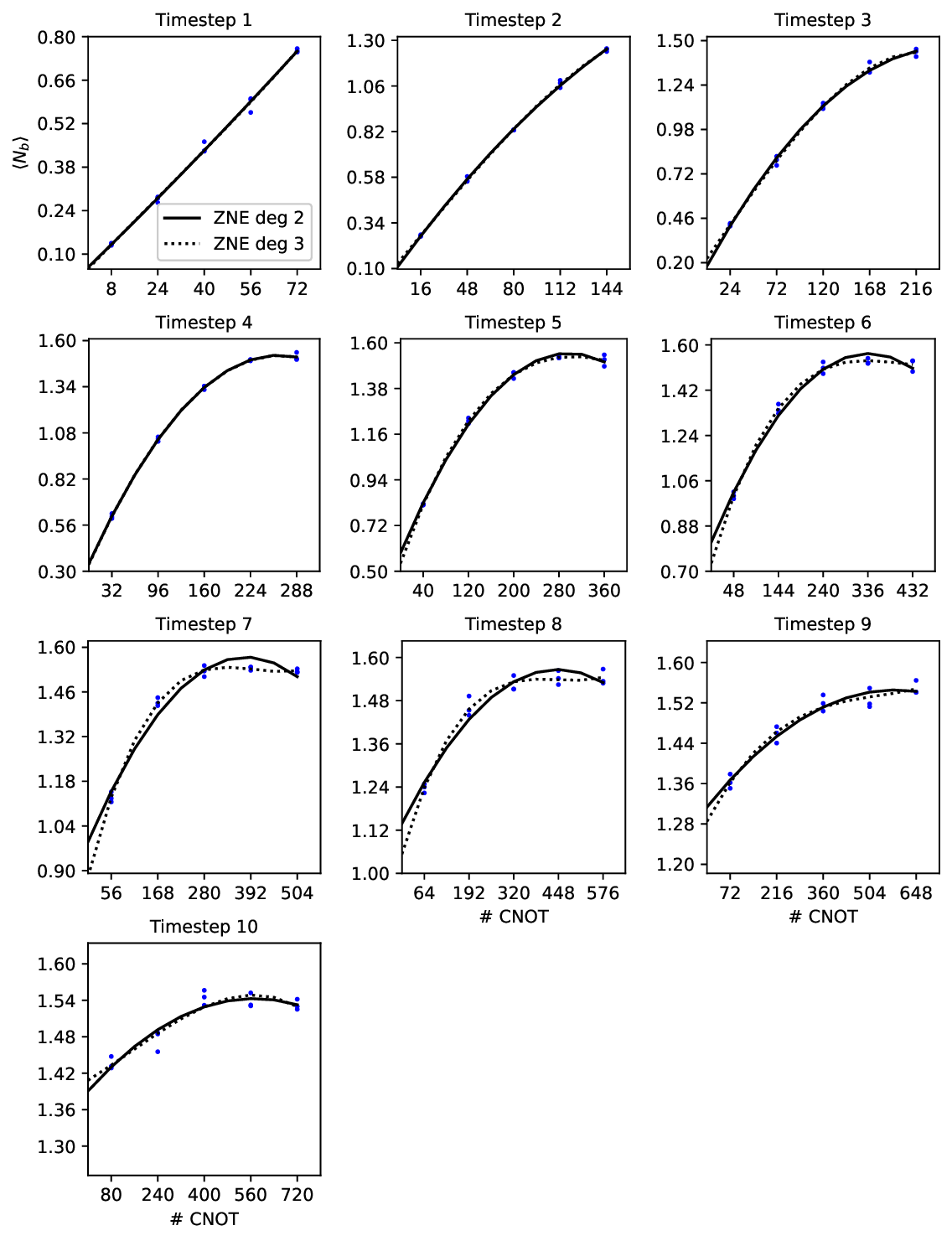}
    \caption{ZNE for the static truncation using three qubits in all time steps. We plot the measured bosonic occupation ( $y$-axis) vs.~ the number of CNOT-gates per circuit ($x$-axis). We repeated each experiment three times (blue dots). We use the data to extrapolate to the 
    noiseless limit (solid line for second order extrapolation, dashed line for third order). }
    \label{fig:zne_static}
\end{figure}

\begin{figure}
    \centering
    \includegraphics[width=1.0\columnwidth]{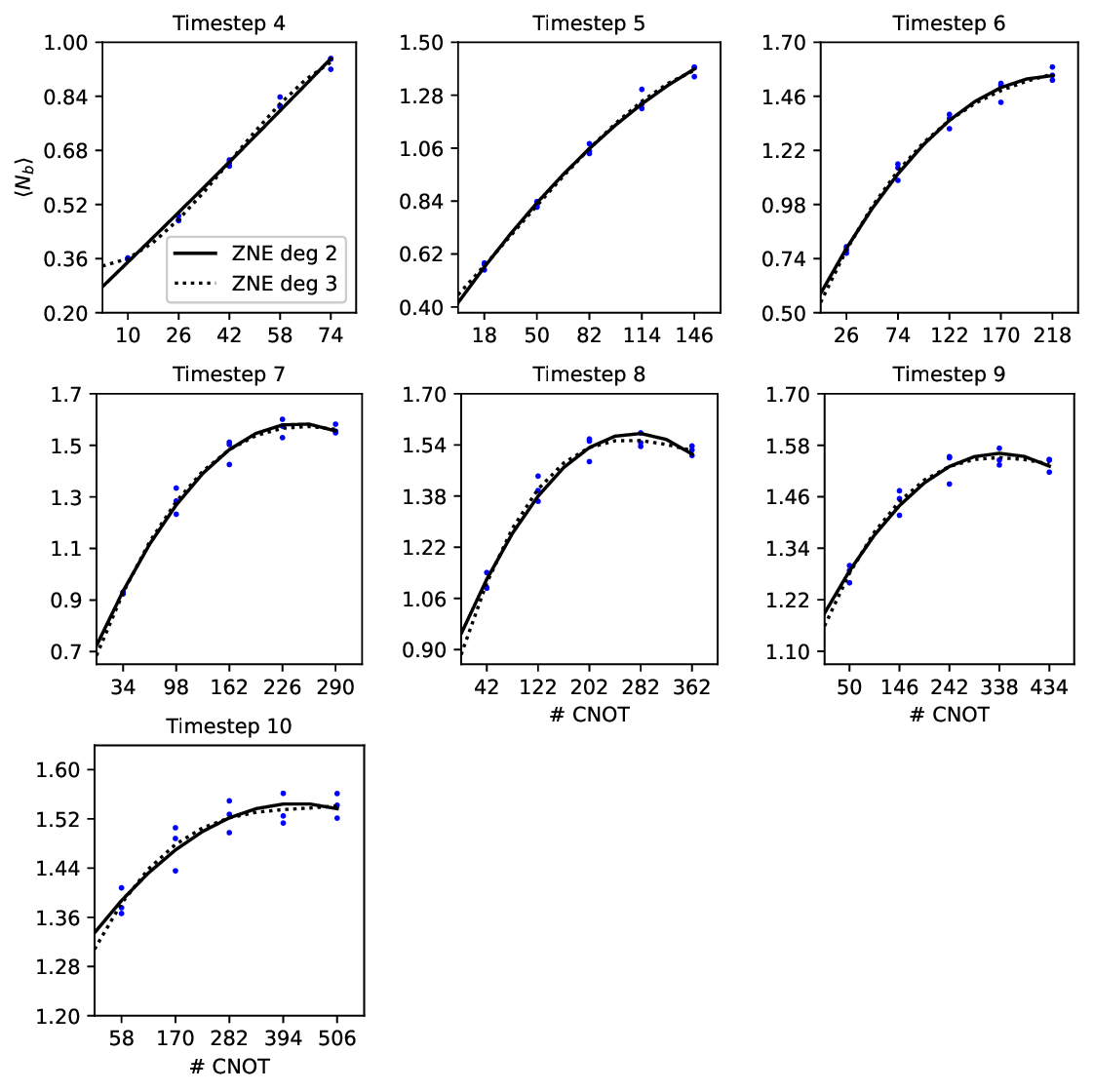}
    \caption{ZNE for the dynamic truncation. We plot the measured bosonic occupation ( $y$-axis) vs.~ the number of CNOT-gates per circuit ($x$-axis). Each experiment is repeated three times (blue dots). We use the data to extrapolate to the zero-noise limit (solid line for second-order extrapolation, dashed line for third-order). The first three time steps are performed using a two-qubit circuit, for which ZNE is not necessary. Thus, the data for these time steps is not shown here.}
    \label{fig:zne_dynamic}
\end{figure}

As derived in Section~\ref{sec:Truncation}, the truncation error of the two-qubit simulation (i.e., one bosonic qubit) exceeds the accuracy-threshold of 10\% during the simulated time interval (cf.~Figure~\ref{fig:TruncationMap}).
However, as also shown in this figure, a truncation with two bosonic qubits is sufficient to simulate the system with a fidelity of 90\% in the time interval of interest. 

We now proceed to extend our two-qubit time-evolution circuit to a three-qubit system with two bosonic qubits.
We compare the results of dynamic and static truncation, allowing up to three bosons exchanged, in Figure~\ref{fig:3Q_Occupation}. As can be inferred from this plot, the dynamic truncation approach offers a significant improvement compared to the naive static truncation for the raw data: the first three data points, for which we used the two-qubit circuits, are -- as expected -- in a good agreement with the theory curve. From step 4 onwards we observe a large deviation, similar to what happens in the static truncation case from the beginning of the evolution.

Even with the dynamic truncation we find a significant deviation between the theory curve and the data simulated by the quantum computer for larger time steps. In order to improve our results and reduce the error we use the ZNE technique [cf. Section~\ref{ZNEbasics}]. 
To perform the extrapolation we measure five data points for different noise levels ($\lambda=1,3\dotsc 9$) and ran each experiment three times.

We perform the ZNE using second- and third-order polynomial extrapolation by performing a linear least-square fit to the data and compare them in Figure~\ref{fig:3Q_Occupation}. The error bars are given by the standard deviation of the intercept that we extract from the fit.
For the static truncation we find that for the first three time steps the second-order extrapolation performs better. Then, from step four onwards it is beneficial to use the third-order extrapolation. For the static truncation we observe the same behaviour for small time steps. However, for larger time steps the difference between both extrapolations is negligible.

In Figures~\ref{fig:zne_static} and \ref{fig:zne_dynamic} we plot the number of CNOT gates vs.~the measured bosonic occupation number, which was used to perform the ZNE. 
Comparing the second- and third-order extrapolations for the first timestep in Figure~\ref{fig:zne_dynamic} one can see that for small circuit noise, it can be beneficial to use the second-order extrapolation since the higher order extrapolation leads to an overfit of the data.
Furthermore, an increase in the number of CNOT gates leads to an increase of the occupation number due to more noise in the quantum circuit. At some point (more precisely, around a total number of $200$ CNOT gates) the curve saturates at a value of $\braket{N_b} = 1.5$. This is expected, since in the three-qubit case 
we have [cf. Equation~\eqref{eq:BosonPartNumber}]
\begin{equation}
\braket{N_b} = 1-\braket{Z_1}+ \frac{1-\braket{Z_0}}{2} \:.
\end{equation}
For a completely depolarized quantum device, we have $\braket{Z_i} = 0.5$ and thus $\braket{N_b}=1.5$. 

\subsection{Adiabatic preparation of the ground- and first-excited states}\label{sec:AQC}
In this section, we apply the time-evolution circuits to eigenstates of the free Hamiltonian $H_0$ to drive them adiabatically to eigenstates of the interacting theory, described by $H=H_0+H_\text{int}$ [cf. Equations~\eqref{ssHint} and \eqref{ssH0}]. 
Adiabatic state preparation is based on the adiabatic theorem of quantum mechanics~\cite{Fock}: a system which is in an eigenstate of a time-dependent Hamiltonian $H(t)=H_0+H_\text{int}(t)$ at an initial time $t_0$, such that its time evolution is governed by the Schrödinger equation
\begin{equation}
    i \frac{\partial}{\partial t} 
 \ket{\psi(t)}=H(t) \ket{\psi(t)},
\end{equation}
will remain an eigenstate, provided that the change of $H(t)$ is ``sufficiently slow.'' (For an up-to-date review in the quantum-computing context, see Ref.~\cite{Albash+Lidar:18}).

\begin{figure}[t]
    \centering
    \includegraphics[width=0.55\columnwidth]{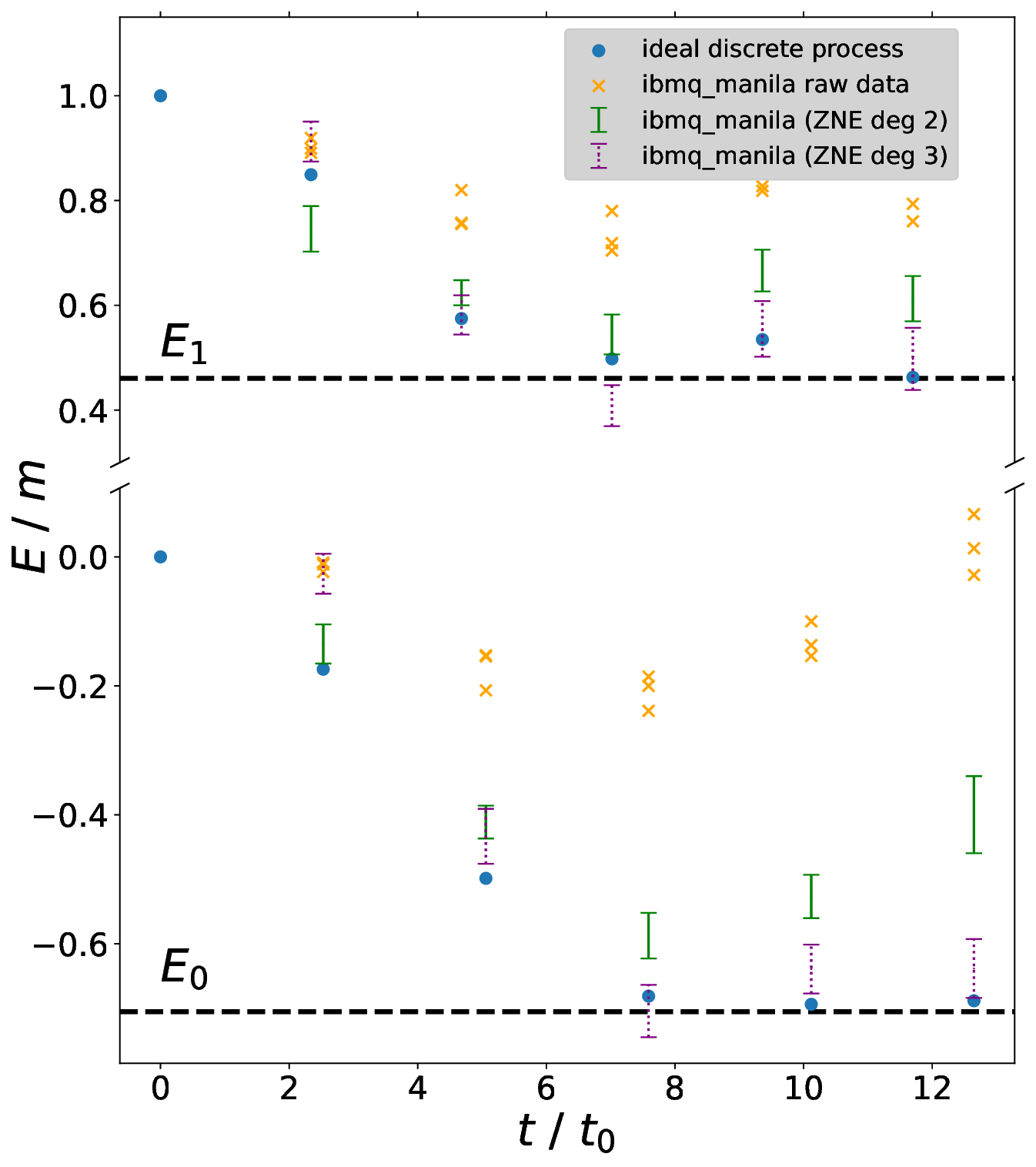}
    \caption{Energy-to-mass ratio $E~/~m$ as a function of dimensionless time $t~/~t_0$ during the adiabatic processes preparing the ground state and first excited state of the interacting theory. We compare second- and third-order extrapolation.}
    \label{fig:EnergyBenchmarks}
\end{figure}

\begin{figure}
    \centering
    \includegraphics[width=1.0\columnwidth]{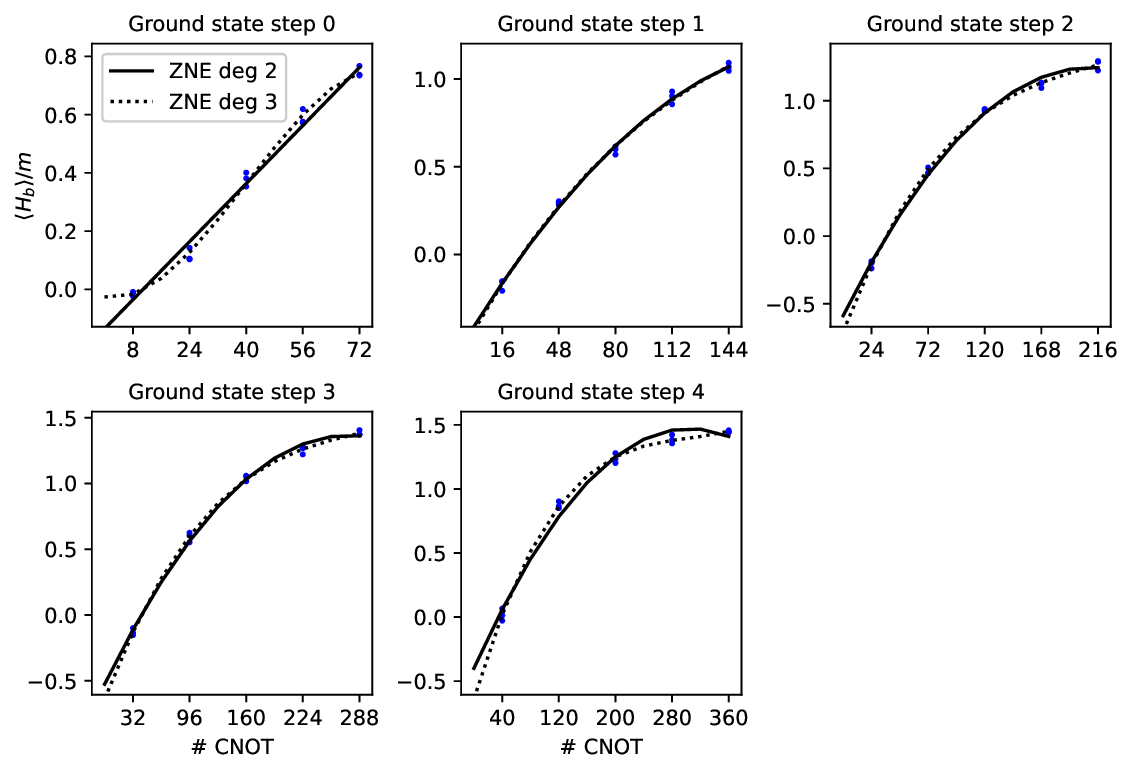}\vspace{0.2 cm}
    \includegraphics[width=1.0\columnwidth]{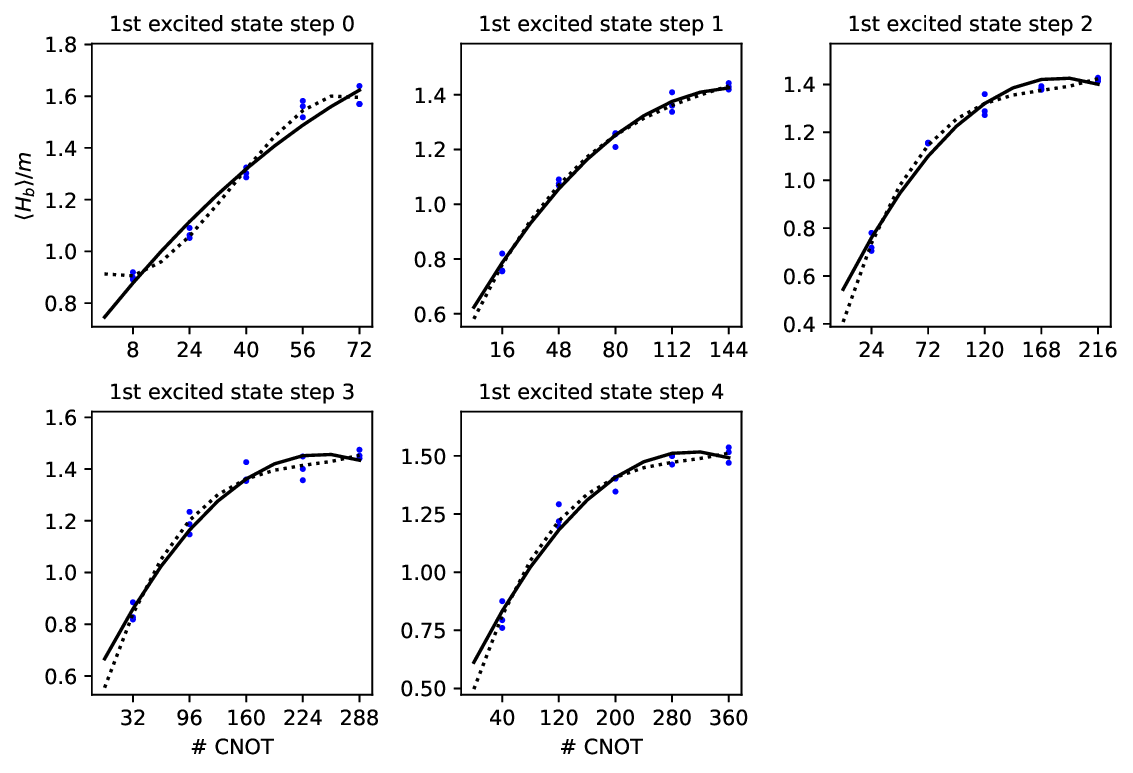}
\caption{ZNE for measuring the energies of the ground state (upper panel) and the first-excited state (lower panel). Each experiment is repeated three times (blue dots). By adding more CNOT gates into the circuit, the expectation value of $H_b/m$ saturates at the value $1.5$. We use second- and third-order extrapolation (solid line for second-order extrapolation, dashed for third-order).}
    \label{fig:ZNE_ground_state}
\end{figure}

For our experiments we made use of the fact that the fermionic particle number is conserved during the whole adiabatic process. Thus, we do not need to measure the fermionic qubit, although we keep it in the circuit as an additional source of noise like before. The bosonic Hamiltonian, for which we consider the adiabatic transition is thus given by [cf.~Equation~\eqref{eq:ThreeQubitHamiltonian2}]:
\begin{align}
    H_b =& -\frac{m}{2}\; ZI - m \; IZ\nonumber \\
    &- \frac{\eta}{2}\left[\frac{1+\sqrt{3}}{2} XI + \frac{1-\sqrt{3}}{2}XZ+\frac{1}{\sqrt{2}}(XX+YY)\right]+\frac{3}{2}m,
\end{align}
and we consider the full Hamiltonian $H=H_b\otimes \mathbbm{1}_f$, where $\mathbbm{1}_f$ denotes the identity operator acting on the fermionic qubit.

In order to find the ground- and first excited state of the interacting single-site Yukawa theory, we start from the corresponding eigenstates of $H_0$ in Equation~\eqref{ssH0}. These states can straightforwardly be prepared as $H_0$ is already diagonal. For instance, to prepare the state with one boson, we simply flip the corresponding qubit in the bosonic register using an $X$-gate. 

Once the desired eigenstate of $H_0$ is prepared, we drive an adiabatic transition using the same quantum circuit as for the time evolution in the three-qubit case (Figure~\ref{fig:ThreeBosonCircuit}), the only difference being that we increase the coupling strength $\eta$ with each time step. 

Ideally, one would choose a large time interval divided into a large number of small steps in order to steadily increase coupling. However, as evident from the benchmarks of the state fidelity in Figure~\ref{fig:3q_fidelity}, the state fidelity decreases significantly with each time step. Hence, we aim to perform the adiabatic transition using only 5 time steps, expecting to keep the fidelity of the final state on \texttt{ibmq\textunderscore manila} above 70\%. 
At the same time, the length of a time step is limited due to the Trotter error. Consequently, we make use of classical computation to determine the optimal time step $\Delta t$, taking the Trotter error into account, to achieve the best possible final state within five steps. 

The optimised discrete processes for both states of interest are illustrated in Figure~\ref{fig:EnergyBenchmarks}. We find that a total simulation time of roughly $t\approx 13\, t_0$ is sufficient to drive the system adiabatically into the eigenstates of the interacting theory. Discretizing that time into five time steps of the length $\Delta t=2.53~t_0$ (ground state) and $\Delta t=2.34~t_0$ (first excited state), suffices to achieve fidelities of the final state of about 99\% for both states under consideration (blue dots in Figure~\ref{fig:EnergyBenchmarks}).

From the calculation of the state fidelities on \texttt{ibmq\textunderscore manila} in the previous section we know that even only these five time steps are too many to produce meaningful results due to the accumulated gate errors. For this reason, we again employ the digital ZNE protocol to reduce the errors of the quantum circuits. 
In Figure~\ref{fig:ZNE_ground_state} we show the data accumulation in order to perform the ZNE. Like before, we repeat each experiment three times and compare second-order and third-order polynomial extrapolation to extract the zero-noise limit of the bosonic Hamiltonian. 
As before, we find that for small noise (first time step) the third-order extrapolation overfits the data [cf.~Figure~\ref{fig:ZNE_ground_state}].
However, as can be seen in Figure \ref{fig:EnergyBenchmarks}, for larger time steps the third-order extrapolation performs better than the second-order extrapolation.

As in the case of the bosonic expectation value in Section~\ref{BosonOccupation} we find an increase in the expectation value in the presence of additional noise. The energy saturates at an expectation value of $\braket{H}_b/m=1.5$ after roughly $200$ CNOT gates. This is expected, since the three-qubit Hamiltonian has a constant term $H_b\propto 1.5 m$.

\section{Summary and Conclusions}  \label{SummConcl}
In this paper we investigated the non-equilibrium dynamics following a quench of Yukawa-type interaction and calculated the two lowest energies of the spectrum of the interaction Hamiltonian within the framework of digital quantum simulation. We showed that a single-site abstraction of scalar Yukawa coupling can successfully be simulated on the existing \texttt{IBMQ} hardware in the regime of low boson-number exchange. 

Using advanced circuit-optimization techniques -- exemplified by the Kraus-Cirac decomposition -- we designed low-depth quantum circuits for simulating the exchange of up to three bosons. In particular, in the one-boson case we designed a constant-depth circuit with only two CNOT gates. This constitutes an example of circuit compression, which is only possible for certain special 
types of the underlying system Hamiltonian. In the three-boson case -- where such a compression is not possible -- we designed a circuit in which one Trotter step requires $8$ CNOT gates, which is far below 
the maximal CNOT-cost of an arbitrary three-qubit gate ($14$ CNOTs)~\cite{Shende+:04}. Finally, by making use of an analogy with the travelling-salesman problem, we also derived a CNOT-cost estimate for quantum circuits emulating the system dynamics for higher boson-number truncations. 

Based on the designed circuits, we characterized the dynamics inherent to the model at hand by computing the expected boson numbers at an arbitrary time after the quench. In addition, by performing state reconstruction using 
quantum-state-tomography algorithms we recovered the Loschmidt echo, a quantity that characterizes the survival probability of the initial vacuum state of the system. Finally, we successfully reproduced an adiabatic process to compute the two lowest eigenvalues of the interacting theory. Using the zero-noise extrapolation technique of error mitigation, we demonstrated a good agreement of our obtained results with numerically-exact classical benchmarks. Our study could in the future be extended to multi-site realizations; this will require much larger quantum processors, ideally with a smaller two-qubit gate error and all-to-all connectivity~\cite{Quantinuum}. 

\vspace{0.35cm}

\noindent {\bf Supplementary Materials:} The following supporting information 
can be downloaded at: https://www.mdpi.com/article/10.3390/quantum6030024/s1, Supplement Code and Data are provided in a supplementary archive.

\vspace{0.175cm}

\noindent {\bf Author Contributions:} Conceptualization, V.M.S.; methodology, T. N. K., V.M.S., M.H., and G. A.; software, T.N.K. and M.H.; validation, T.N.K. and M.H.; formal analysis, V.M.S. and G.A.; investigation, V.M.S.; data curation, T.N.K. and M.H.; writing---original draft preparation, V.M.S.; writing---review and editing, T.N.K., M.H., and G.A. All authors have read and agreed to the published version
of the manuscript.

\vspace{0.175cm}

\noindent {\bf Funding:} This research was supported in part by the Deutsche Forschungsgemeinschaft (DFG) -- SFB 1119 -- 236615297 (G. A. and V. M. S.) and in part by the research project ``Zentrum f\"{u}r Angewandtes Quantencomputing" (ZAQC), which is funded by the Hessian Ministry for Digital Strategy and Innovation and the Hessian Ministry of Higher Education, Research and the Arts (T. N. K. and M. H.).

\vspace{0.175cm}

\noindent {\bf Data Availability Statement:} The raw data supporting the conclusions of this article will be made available by the authors on request.

\vspace{0.175cm}

\noindent {\bf Acknowledgments:} We would like to acknowledge useful discussions on the field-theoretical description of Yukawa coupling with H.-W. Hammer. We acknowledge the use of IBM Quantum services for this work. The views expressed are those of the authors, and do not reflect the official policy or position of IBM or the IBM Quantum team.

\vspace{0.175cm}

\noindent {\bf Conflicts of Interest:} The authors declare no conflicts of interest.

\appendix 
\section[\appendixname~\thesection]{Derivation of the single-site Hamiltonian} \label{DeriveHamiltonian}
The free Hamiltonians of the Dirac- and Klein-Gordon fields in momentum space are given by Equation~\eqref{EqFreeHamiltonian}
in the main text. By discretizing to a single grid point with momentum $\textbf{p}=0$, we have
\begin{equation}
H_0 = H_\text{Dirac} + H_\text{KG} = M \sum_s \left(a^{s\dagger} 
a^{s} + c^{s\dagger} c^{s}\right) + m b^{\dagger} b \:.
\end{equation}
By ignoring the spin degrees of freedom (spinless fermions), we find the single-site
version of the free Hamiltonian:
\begin{equation}
H_0 = H_\text{Dirac} + H_\text{KG} = M (a^{\dagger} a + c^{\dagger} c) + m b^{\dagger} b \:.
\end{equation}

In its most general form, the interaction Hamiltonian is given by 
[cf. Equation~\eqref{fullHint} in Section~\ref{scalarYukawa} of the main text]
\begin{align}
H_\text{int} =& \frac{g a_{l}^{3/2}}{2} \sum_{\textbf{p}, \textbf{p}'} \sum_{r, s} \bigg\{ 
\frac{1}{\sqrt{8 \Omega_\textbf{p}\Omega_{\textbf{p}'}\omega_{\textbf{p}-\textbf{p}'}}}  
\left[a_\textbf{p}^{r\dagger} \bar u^r(\textbf{p}) + c_{-\textbf{p}}^{r} \bar v^r(-\textbf{p})\right]\nonumber\\
&\quad\times\left[a_{\textbf{p}'}^s  u^s(\textbf{p}') + c_{-\textbf{p}'}^{s\dagger}  v^s(-\textbf{p}')\right]
\left(b_{\textbf{p}-\textbf{p}'}+ b_{\textbf{p}'-\textbf{p}}^\dagger\right)
+\text{H.c.} \bigg\}\:.
\end{align} 
By restricting momentum space to a single site with $\textbf{p}=0$, we obtain
\begin{align} \label{SingleSiteOrigin}
        H_\text{int} &= \frac{g a^{3/2}}{2\sqrt{8 M^2m}}  \sum_{r, s} \left[
        \left(a^{r\dagger} \bar v^r + c^{r} \bar u^r\right)
        \left(a^s  v^s + c^{s\dagger}  u^s\right)
        \left(b+ b^\dagger\right)
        + \text{H.c.} \right]\nonumber \\
        &= \frac{g a_l^{3/2}}{2\sqrt{8 M^2m}} \sum_{r, s} \left[ 
        \left(
        a^{r\dagger} a^s \bar u^r u^s
        + a^{r\dagger} c^{s\dagger} \bar u^r v^s
        + c^{r} a^s  \bar v^r u^s  
        + c^{r} c^{s\dagger} \bar v^r v^s
        \right)\right.\nonumber\\
        &\left.\qquad\times
        \left(b+ b^\dagger\right)
        + \text{H.c.} \right] \:,
\end{align}
where the momentum index/argument is dropped. For the spinor products 
in the non-relativistic limit we have
\begin{equation}
    \begin{split}
        \bar u^r u^{s'} &= 2M \delta^{rs}\:,    \\
        \bar u^r v^{s'} &= 0 \:,  \\
        \bar v^r u^{s'} &= 0\:,  \\
        \bar v^r v^{s'} &= -2M \delta^{rs}\:.
    \end{split}
\end{equation}
By taking into account these last identities, the single-site Hamiltonian in Equation~\eqref{SingleSiteOrigin}
reduces to 
\begin{equation}
    \begin{split}
        H_\text{int} &= 
        \frac{g a_l^{3/2}}{\sqrt{2m}} \sum_{s} 
            \left(
            a^{s\dagger} a^s 
            - c^{s} c^{s\dagger} 
            \right)
            \left(b+ b^\dagger\right) \\
        &=  \frac{g a_l^{3/2}}{\sqrt{2m}} \sum_{s} 
            \left(
            a^{s\dagger} a^s 
            + c^{s\dagger} c^{s} - 1 
            \right)
            \left(b+ b^\dagger\right) \:.
    \end{split}
\end{equation}

Finally, we ignore the spin degrees of freedom and restrict ourselves to the 
states with the vanishing total charge, which yields
\begin{equation}
H_\text{int} = \frac{g a_l^{3/2}}{\sqrt{2m}}
\left(a^{\dagger} a+c^{\dagger} c -1\right)
\left(b+ b^\dagger\right) \:.
\end{equation}           
At this point we can express the lattice spacing $a_l$ in terms of the boson mass via $a_l/(2m)=\beta\gamma$. 
Because $\beta$ ought to be small for our single-site abstraction (the nonrelativistic limit), we 
use $\gamma\approx 1$. In this manner, we finally obtain the zero-dimensional interaction Hamiltonian
\begin{equation}
H_\text{int} = \frac{\eta}{2}\left(a^{\dagger} a+c^{\dagger} c -1\right)
\left(b+ b^\dagger\right)\:,
\end{equation}
with $\eta\equiv 4mg\beta^{3/2}$ being the effective coupling strength.

\section[\appendixname~\thesection]{Boson mapping} \label{BosonMapping}
The bosonic creation operator with a truncation at $\Lambda = 2^N-1$ can be represented by the matrix
\begin{equation}
    b_N^\dagger = \sum_{k=1}^{2^N-1}\sqrt{k}~G_{N,k},
\end{equation}
where $(G_{N,k})_{ij} = \delta_{i-1,k}\delta_{j,k}$. By defining $I_\pm = (I\pm Z)/2$ and 
$\sigma_\pm = (X\pm iY)/2$, we derive a recurrence relation for $G_{N,k}$ in terms of the qubit number $N$:
\begin{equation}
    G_{N+1, k} = 
    \begin{cases}
        I_+\otimes G_{N,k} & \text{if } k \leq 2^N-1, \\
        \sigma_- \otimes \sigma_+^{\otimes N} & \text{if } k=2^N, \\
        I_- \otimes G_{N, k-2^N} & \text{if } k > 2^N.
    \end{cases}
    \label{eq:BosonCreationRec}
\end{equation}

By representing $k$ as a bit string [cf. Equation~\eqref{BinaryRepresent} in Section~\ref{EncodeBosons}] 
we can straightforwardly find the explicit solution to Equation~\eqref{eq:BosonCreationRec}:
\begin{equation}
G_{N,k} = \left(\frac{1}{2}\right)^N\bigotimes_{j=0}^{N-1} F_{j,k}.
\end{equation}

The single-qubit operators $F_{j,k}$ depend on the binary string $q(k)$ as follows:
\begin{equation}
    F_{j,k} =
    \begin{cases}
        I_j + Z_j   & \text{if}~ q(k)_j = 0   ~\text{and}~ \exists~m < j: q(k)_m = 1, \\
        I_j - Z_j   & \text{if}~ q(k)_j = 1   ~\text{and}~ \exists~m < j: q(k)_m = 1, \\
        X_j + i Y_j & \text{if}~ q(k)_j=0     ~\text{and}~ \forall~m < j: q(k)_m = 0, \\
        X_j - i Y_j & \text{if}~ q(k)_j=1     ~\text{and}~ \forall~m < j: q(k)_m = 0, \\
    \end{cases}
\end{equation}
which can be written more compactly as
\begin{equation}
    F_{j,k} =
    \begin{cases}
        I_j + (-1)^{q_j(k)} Z_j   & \text{if}~ \exists~m < j: q(k)_m = 1, \\
        X_j + (-1)^{q_j(k)} i Y_j & \text{if}~ \forall~m < j: q(k)_m = 0. \\
    \end{cases}
\end{equation}

\section[\appendixname~\thesection]{Kraus-Cirac decomposition} \label{KrausCiracDecomp}
In this section, we show how the unitary gate $\mathcal{S}$, defined in Equation~\eqref{defineS}, is decomposed 
into a quantum circuit. 

Every special unitary operation $U\in \text{SU}(4)$ can be written as
\begin{equation}
    U = (K_4 \otimes K_3) \: N(\alpha, \beta, \gamma) \: \left(K_2 \otimes K_1\right),
    \label{eq:KrausCiracDecomposition}
\end{equation}
where $K_1,\:K_2,\:K_3,\:K_4  \in \text{SU}(2)$ and
\begin{equation}
    N(\alpha, \beta, \gamma) = \exp\left[i\left(\alpha XX + \beta YY + \gamma ZZ\right)\right],
\end{equation}
with $\alpha, \beta, \gamma \in \mathbb{R}$. Assuming that $\mathcal{S}$ takes the form from 
\eqref{eq:KrausCiracDecomposition}, we wish to find the single qubit gates $K_j$ and the two-qubit 
rotation angles $\alpha, \beta, \gamma$. We start by transforming $\mathcal{S}$ to the magic basis. 
By making use of the magic gate
\begin{equation}
\mathcal{M} = \frac{1}{\sqrt{2}} 
\begin{pmatrix}
        1 & i & 0 & 0 \\
        0 & 0 & i & 1 \\
        0 & 0 & i & -1 \\
        1 & -i & 0 & 0 \\
\end{pmatrix}\:,
\end{equation}
we find
\begin{align}
&V = \mathcal{M}^\dagger \mathcal{S} \mathcal{M} \nonumber\\
  &= \frac{1}{4} 
\begin{pmatrix}
 -(\tilde\lambda_-+\tilde\lambda_+-2) & -i(\tilde\lambda_--\tilde\lambda_+)  
 & i(\tilde\lambda_-+\tilde\lambda_++2)  & \tilde\lambda_--\tilde\lambda_+\\
 i(\tilde\lambda_-+\tilde\lambda_++2) & -(\tilde\lambda_- - \tilde\lambda_+)    
 & \tilde\lambda_-+\tilde\lambda_+-2     & -i(\tilde\lambda_--\tilde\lambda_+) \\
 -i(\tilde\lambda_--\tilde\lambda_+)     & -(\tilde\lambda_-+\tilde\lambda_+-2)    
 & -(\tilde\lambda_--\tilde\lambda_+)      & -i(\tilde\lambda_-+\tilde\lambda_++2) \\
 -(\tilde\lambda_--\tilde\lambda_+)        & i(\tilde\lambda_-+\tilde\lambda_++2)  
 & i(\tilde\lambda_--\tilde\lambda_+)    & -(\tilde\lambda_-+\tilde\lambda_+-2) \\
\end{pmatrix}
\end{align}

The transformation to the magic basis has two beneficial properties. First, it diagonalizes 
$N(\alpha,\beta,\gamma) = \mathcal{M}D\mathcal{M}^\dagger$ such that
\begin{equation}
D = \text{diag}~\left(e^{i(\alpha-\beta+\gamma)}, 
e^{-i(\alpha-\beta-\gamma)}, e^{i(\alpha+\beta-\gamma)}, 
e^{-i(\alpha+\beta+\gamma)} \right).
\end{equation}

Second, the mapping $A\otimes B \to \mathcal{M}^\dagger (A\otimes B) \mathcal{M}$ precisely defines 
the isomorphism between the spaces $\text{SU}(2)\otimes \text{SU}(2)$ and $\text{SO}(4)$. Consequently,
 we may write $V = Q_1 D Q_2^T$, where the special orthogonal matrices $Q_1, Q_2 \in \text{SO}(4)$ and 
the diagonal matrix $D$ are obtained by performing a singular value decompositon. Here, we construct 
$V^TV = Q_2 D^2 Q_2^T$ with
\begin{equation}
    V^T V = \frac{1}{2}
    \begin{pmatrix}
        -(a+b)  & -i(a-b)   & 0         & 0 \\
        -i(a-b) & -(a+b)    & 0         & 0 \\
        0       & 0         & -(a+b)    & i(a-b) \\
        0       & 0         & i(a-b)    & -(a+b) \\
    \end{pmatrix}
\end{equation}

Through diagonalization, which can be performed block-wise here, we find
\begin{equation}
    Q_2 = \frac{1}{\sqrt{2}} 
    \begin{pmatrix}
        0 & 0 & 1 & 1 \\
        -1 & 1 & 0 & 0 \\
        0 & 0 & -1 & 1 \\
        1 & 1 & 0 & 0 \\
    \end{pmatrix},
\end{equation}
and
\begin{equation}
    D = \text{diag}~\left(e^{i\varphi}, e^{i\varphi}, e^{-i\varphi}, e^{-i\varphi}\right),
\end{equation}
with $\varphi = \arctan[\sqrt{2}/(1+\sqrt{3})]$. Thus, we 
conclude that $\alpha=\beta = 0$ and $\gamma = \varphi$. $Q_1$ is straightforwardly obtained through $Q_1 = V Q_2 D^{-1}$. The rest of the derivation, which will not be shown here, is rather straightforward as it only entails the inverse basis transformation (i.e. the one from the magic basis to the original one). 

\section[\appendixname~\thesection]{The distance metric} \label{DistanceMetric}
In this section, we derive the CNOT-cost of a time-step using the star+ancilla layout, depicted in Figure~\ref{fig:ZZZcircuits}(b). 
While the pure star layout from Figure~\ref{fig:ZZZcircuits}(a) makes use of a smaller number of CNOT gates, the ancilla qubit allows for easier gate cancellation in the transition zones, as no basis transformations are applied on the target qubit. 

Let us consider the transition zone of two consecutive Pauli strings $P_1$ and $P_2$. We will briefly describe the approach of Ref.~\cite{Gui+:20} before introducing our improved cancellation technique. 
On indices where $P_1[i] = P_2[i]$, the CNOT gates cancel out. But for indices where $P_1[i] \neq P_2[i]$, one has to differentiate between two cases: (a) neither $P_1[i]$ nor $P_2[i]$ is $I$ or (b) one of them is $I$. 
Case (a) requires two CNOTs, while (b) only needs one. Combining these results, the CNOT distance metric of Ref.~\cite{Gui+:20} (cf.~Equation~(15) in that reference) is defined as
\begin{equation}
|P_1-P_2|_\text{CNOT} := \sum_{i \in [N]} 1_{P_1[i]
\neq P_2[i]}\left(1+ 1_{I \notin \{P_1[i], P_2[i]\}}\right).
    \label{eq:CNOTDist}
\end{equation}

\begin{figure}[h!]
\centering
\includegraphics[width=0.7\textwidth]{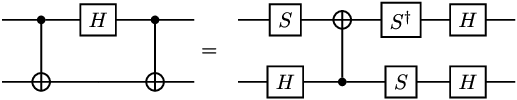}
\caption{\label{fig:XZ}CNOT reduction in the $X\leftrightarrow Z$ transition circuit.}
\end{figure}

We will now modify this cost function by using circuit identities which reduce the CNOT-cost in case (a) to one. We start by considering transitions of the type 
$X \to Z$ or $Z \to X$, represented by the circuit in Figure~\ref{fig:XZ}, which can be reduced to a single 
CNOT gate.

\begin{figure}[h!]
\centering
\includegraphics[width=\textwidth]{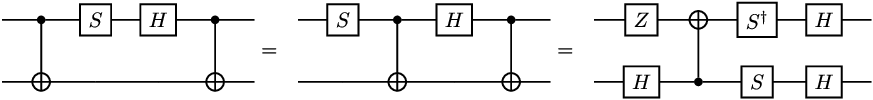}
\caption{\label{fig:ZY}CNOT reduction in the $Z \to Y$ transition circuit.}
\end{figure}

Next, we investigate the $Z \to Y$ and $Y \to Z$ transitions. Of course, 
$Y \to Z$ is just the inverse of $Z \to Y$ and can be obtained as the Hermitian conjugate. 
The circuit for $Z \to Y$ is shown in Figure~\ref{fig:ZY}. In the first step, we used that $S$ 
commutes with a control, and in the second one, we reuse the circuit from Figure~\ref{fig:XZ} 
for the $X\to Z$ transition.

\begin{figure}[h!]
\centering
\includegraphics[width=\textwidth]{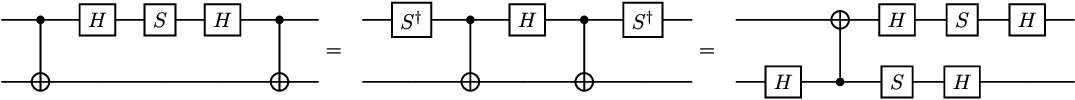}
\caption{\label{fig:XY}CNOT reduction in the $X \to Y$ transition circuit.}
\end{figure}

It remains to simplify the circuit for $X\to Y$, depicted in Figure~\ref{fig:XY}. 
In the first step, we use that $HSH = S^\dagger H S^\dagger$ up to a global phase 
and the commutation of $S^\dagger$ with the control. In the second step, we once 
again recycle the circuit for $X\to Z$ and undo the global phase.

All these results together prove that any transition with $P_1[i]\neq P_2[i]$ 
can be implemented with only one CNOT gate, which implies that we have
\begin{equation}
|P_1-P_2|_\text{CNOT} := \sum_{i \in [N]} 1_{P_1[i]
\neq P_2[i]} = |P_1-P_2|_\text{Ham}  \:,
\end{equation}
where $|P_1-P_2|_\text{Ham}$ is the Hamming distance counting the number of disagreeing indices. 
Compared to the previous cost function in Equation~\eqref{eq:CNOTDist} this cuts the cost by up to $50\%$. 

\section[\appendixname~\thesection]{Generation of Pauli strings} \label{GeneratePauliStrings}
In this appendix, we derive an algorithm to efficiently calculate the Pauli strings for a truncation 
using $N$ qubits. Our approach is based on finding a recurrence relation between $b+b^\dagger$ truncated 
with $N$ and $N+1$ qubits~\cite{Huang+:21}. Writing out $b+b^\dagger$ in the Fock state basis, we find:
\tiny
\begin{equation*}
\left(
\begin{array}{ccccc|cccccc} 
  0 & \sqrt{1} & \cdots & 0 & 0   \\ 
  \sqrt{1} & 0 & \cdots & 0 & 0 \\
  \vdots & \vdots & \ddots & \vdots & \vdots \\
  0 & 0 & \cdots & 0 & \sqrt{2^N-1} \\
  0 & 0 & \cdots & \sqrt{2^N-1} & 0 & \sqrt{2^N} \\ \hline
  & & & & \sqrt{2^N} & 0 & \sqrt{2^N+1}  & \cdots & 0 & 0   \\ 
  & & & & & \sqrt{2^N+1} & 0 & \cdots & 0 & 0 \\
  & & & & & \vdots & \vdots & \ddots & \vdots & \vdots \\
  & & & & & 0 & 0  & \cdots & 0 & \sqrt{2^{N+1}-1}  \\
  & & & & & 0 & 0 & \cdots & \sqrt{2^{N+1}-1} & 0
\end{array}
\right)\:. 
\end{equation*}
\normalsize
Note that the top left block is nothing but $(b+b^\dagger)_N$. In order to succinctly express the recurrence relation, we define $I_\pm = (I\pm Z)/2$ and $\sigma_\pm = (X\pm iY)/2$. The expression we find is given by
\begin{equation}
(b+b^\dagger)_{N+1} = (b+b^\dagger)_{N} \otimes I_+ + 2^N \left(\sigma_+^{\otimes N}
\otimes\sigma_-  + \sigma_-^{\otimes N}\otimes \sigma_+\right) + M_{\text{BR}, N} 
\otimes I_-  \:,
\end{equation}
where $(b+b^\dagger)_1 = X$ and $M_{\text{BR}, N}$ is the bottom right sub-matrix of the matrix above. 
The reason we do not write it out explicitly is that $M_{\text{BR}, N}$ consists of the same Pauli 
strings as $(b+b^\dagger)_{N}$. Bearing this in mind, we write down the recurrence relation 
\begin{equation}
S_{N+1} =  S_N \otimes I \cup S_N \otimes Z \cup S(\sigma_+^{\otimes N}\otimes\sigma_-  
+ \sigma_-^{\otimes N}\otimes \sigma_+)  \:,
\end{equation}
for the set of Pauli strings $S_N$ [cf. Equation~\eqref{eq:PauliStringsRec}], with $S_1 = \{X\}$.

Here, the Pauli operators $I$ and $Z$ come from the expansion of $I_\pm$. As discussed in 
Section~\ref{CNOTcostHigherTrunc}, the set $S(\sigma_+^{\otimes N}\otimes\sigma_-  + \sigma_-^{\otimes N}
\otimes \sigma_+)$ consists of $2^N$ Pauli strings, which implies that the number of Pauli 
strings $|S_N|$ satisfies the difference equation
\begin{equation}
|S_{N+1}| = 2 |S_N| + 2^N \:,
\end{equation}  
with $|S_1| = 1$. The explicit solution to this equation is given by
\begin{equation}
|S_N| = N~2^{N-1} \:.
\end{equation}

\section{Upper bound on the CNOT-cost} \label{CNOTcostBound}
Using the recurrence formula for the Pauli strings \eqref{eq:PauliStringsRec} and the functions for the 
CNOT-cost of the start- and end layer as well as the transition zones, we will now derive an upper bound 
on the CNOT-cost of the circuit representing $\exp\left(-i H_\text{b}\Delta t \right)$ in first-order 
Trotterization. We first recall the recurrence formula for the Pauli strings [cf. Equation~\eqref{eq:PauliStringsRec}] 
\begin{equation}
S_{N+1} =  S_N \otimes I \cup S_N \otimes Z \cup S(\sigma_+^{\otimes N}\otimes\sigma_-  
+ \sigma_-^{\otimes N}\otimes \sigma_+)  \:,
\end{equation}
where $S_1 = \{X\}$. 

In the following, $C(O)$ denotes the CNOT-cost of implementing the Pauli strings of the operator $O$
in first-order Trotterization. Using the identities $C(S_N \otimes I) = C(S_N)$ and $C(S_N \otimes Z) = C(S_N) + 2$,
as well as the fact that we can always cancel $2N$ CNOT-gates when transitioning from $S_N \otimes I$ to 
$S_N \otimes Z$, we find
\begin{equation}
C(S_N \otimes I \cup S_N \otimes Z) 
= 2 C(S_N) + 2 - 2N \:.
\end{equation}

For the second part, namely $S(\sigma_+^{\otimes N}\otimes\sigma_-  + \sigma_-^{\otimes N}\otimes 
\sigma_+)$, we recall that this set contains all $(N+1)$-digit Pauli strings made from $X$ and $Y$ 
with an even number of $Y$s. The start- and end layers of this set give us $2(N+1)$ CNOTs. To maximize 
the gate cancellation, we arrange the set using every other element of an $(N+1)$-digit gray code. 
This way, only two characters change per transition. For $2^N-1$ transition zones, we then find a 
total cost of $2^{N+1}-2$. Combining these results, we arrive at the difference equation
\begin{equation}
C(S_{N+1}) \leq 2 C(S_N) + 2^{N+1} \:,
\end{equation}
with $C(S_1) = 2$. The explicit solution to this equation is given by
\begin{equation}
C(S_N) \leq N~2^N = 2 |S_N| \:.
\end{equation}

\end{document}